\documentclass[fleqn,usenatbib]{mnras}

\usepackage{newtxtext,newtxmath}

\usepackage[T1]{fontenc}

\DeclareRobustCommand{\VAN}[3]{#2}
\let\VANthebibliography\thebibliography
\def\thebibliography{\DeclareRobustCommand{\VAN}[3]{##3}\VANthebibliography}

\usepackage{graphicx}	
\usepackage{amsmath}	
\usepackage{amssymb}	
\usepackage{todonotes}
\usepackage{lipsum}
\usepackage{ulem}
\usepackage{xcolor}

\defcitealias{kaspi}{K13}

\title[Constraining the Depth of Winds on Ice Giants]{Constraining the Depth of the Winds on Uranus and Neptune \newline
via Ohmic Dissipation}
\author[D. Soyuer et al.]{
Deniz Soyuer,$^{1}$\thanks{E-mail: deniz.soyuer@uzh.ch}
Fran\c{c}ois Soubiran,$^{2}$
Ravit Helled$^{1}$
\\
$^{1}$Center for Theoretical Astrophysics and Cosmology, Institute of Computational Science, University of Zurich,
Winterthurerstrasse 190, 8057 Zurich, \\
Switzerland\\
$^{2}$CEA DAM-DIF, 91297 Arpajon, France
}

\date{Accepted 2020 August 10. Received 2020 August 5; in original form 2020 June 18.}

\pubyear{2020}

\begin{document}
\label{firstpage}
\pagerange{\pageref{firstpage}--\pageref{lastpage}}
\maketitle

\begin{abstract} Determining the depth of atmospheric winds in the outer planets of the Solar System is a key topic in planetary science. We provide constraints on these depths in Uranus and Neptune via the total induced Ohmic dissipation, due to the interaction of the zonal flows and the planetary magnetic fields. An upper bound can be placed on the induced dissipation via energy and entropy flux throughout the interior. The induced Ohmic dissipation is directly linked  to the electrical conductivity profile of the materials involved in the flow. We present a  method for calculating electrical conductivity profiles of ionically conducting hydrogen--helium--water mixtures under planetary conditions, using results from \textit{ab initio} simulations. We apply this prescription on several ice giant interior structure models available in the literature, where all the heavy elements  are represented by water. According to the energy (entropy) flux budget, the maximum penetration depth for Uranus lies above $0.93  R_{\mathrm{\scriptscriptstyle{U}}}$ ($0.90  R_{\mathrm{\scriptscriptstyle{U}}}$) and for Neptune  above  $0.95 R_{\mathrm{\scriptscriptstyle{N}}}$ ($0.92 R_{\mathrm{\scriptscriptstyle{N}}}$). These results for the  penetration depths are upper bounds, and are  consistent with previous estimates based on the contribution of the zonal winds to the gravity field.
As expected, interior structure models with higher water abundance in the outer regions have also a higher electrical conductivity and therefore reach the Ohmic limit at shallower regions.  
Thus, our study shows that the likelihood of deep-seated winds on Uranus and Neptune drops significantly with the presence of water in the outer layers.


\end{abstract}

\begin{keywords}
planets and satellites: individual: Uranus, -- planets and satellites: individual: Neptune -- planets and satellites: composition -- planets and satellites: interiors -- planets and satellites: magnetic fields -- methods: data analysis
\end{keywords}



\section{Introduction}
Uranus (U) and Neptune (N) are the outermost planets in the Solar System. They both exhibit strong atmospheric winds, with speeds up to $\sim$200 ms$^{-1}$ and $\sim$400 ms$^{-1}$ in the System III frame \citep{warwick_ura, warwick_nep}, respectively. 
These zonal winds are considered to be symmetric with respect to the equator, with a retrograde motion in central latitudes and prograde in higher latitudes \citep{smith_original, hammel_original}. 
It is however unknown whether the winds extend into the deep layers of the planets or are confined to shallow regions. 
\citet{kaspi} (hereafter \citetalias{kaspi}) have investigated the maximum penetration depth of zonal winds on Uranus and Neptune via gravitational constraints. By estimating the dynamical density contribution of the winds on the gravity harmonic $J_4$, they determined the maximum penetration depth of the winds to be roughly $\sim 0.95R_{\mathrm{\scriptscriptstyle{U,N}}}$ for both Uranus and Neptune. Naturally, these limits depend on the assumed internal structure models. Hence, they have considered a large range of radial density profiles for both planets where the densities were represented by a 6th order polynomial \citep{helled}~above a constant density core with various masses and sizes. 
A similar method has also been applied to the zonal winds of Jupiter  and Saturn, constraining the wind depths from Juno and Cassini gravity data to roughly 0.95$R_{\mathrm{\scriptscriptstyle{J}}}$ and 0.85$R_{\mathrm{\scriptscriptstyle{S}}}$, respectively \citep{kaspi2018, kaspi2019,  galanti,  J_s}.
Since higher order gravity harmonics have a greater contribution to the shape of the density profile in the outermost regions \citep[e.g. Figure 1 in][]{helled} and odd harmonics carry information on density asymmetries, this formulation gives more consistent results for the gas giants, due to their accurately measured  higher order harmonics and non-zero odd numbered  harmonics $J_{i\ge3}$ measured for Jupiter with the Juno spacecraft and for Saturn with the Cassini spacecraft, compared to that of Uranus and Neptune.

An alternative approach to the penetration depth question has been explored thoroughly for the gas giants, in which the maximum depth of the zonal winds is constrained via the induced Ohmic dissipation due to the interaction between the planetary magnetic field and the zonal flow, where the electrical conductivity of the constituents of the flow are expected to increase with depth. The main idea is to place an upper bound on the induced Ohmic dissipation inside the planet using heat flux balance \citep{liu_phd, liu, cao2017, wicht} and entropy flux balance \citep{jones, wicht}. 
Since both the magnetic field strength and the electrical conductivity increase with depth, the dissipation is expected to increase dramatically as well (potentially outshining the planet). Therefore, one can calculate for a given interior structure model, bulk composition and wind behaviour, whether the total induced Ohmic dissipation is allowed in the heat flux budget (or the entropy flux budget) of the system. The flux limits would then provide an upper bound for the total Ohmic dissipation, hence limiting the maximum penetration depth of the winds for a given penetration model.
 
It is important to keep in mind that these constraints are motivated by a purely kinematic argument, where one assumes the magnetic field is not changing in time,hence its interaction with the zonal flow does not produce significantly strong magnetic fields to break this equilibrium. In other words, the induction equation describing the evolution of the planetary magnetic field is in a quasi-steady-state:
\begin{equation}
\label{eq:ind}
0 \approx \frac{\partial \mathbf{B}}{\partial t} = \nabla \times (\mathbf{U_{\textrm{conv.}}} \times \mathbf{B})
- \nabla \times \left(\frac{1}{\mu_0 \sigma}\nabla \times \mathbf{B}\right). 
\end{equation}
Here, the first term on the r.h.s is the generative component. Note that,  $\mathbf{U_{\textrm{conv.}}}$ describes the (relatively slow) convective flows in the dynamo region, generating the planetary magnetic field, predicted to lie at depths $0.7 - 0.8$ $R_{\mathrm{\scriptscriptstyle U,N}}$ in the ice giants \citep{stanley, redmer}. 
The second term describes the dissipative component, with $\mu_0$ the vacuum permeability and $\sigma$ the electrical conductivity of the materials involved. Usually the product is written in terms of the magnetic diffusivity: $\eta = (\mu_0\sigma)^{-1}$.

An important parameter when describing magnetic field generation in convective environments is the magnetic Reynolds number which is  given by: $R_\textrm{m} =UL/\eta$, where $U$ and $L$ are the typical speed and the length scales of the system. $R_\textrm{m}$ is a dimensionless ratio of the generative and the dissipative terms in Eq.~(\ref{eq:ind}). Interesting cases are the diffusive limit; described by $R_\textrm{m} \ll 1$, in which the magnetic field lines are relaxed and are not locked in with the flow, and the inductive limit; $R_\textrm{m} \gg 1$, where the magnetic field lines are dragged along with the flow \citep[i.e. Alfv\'{e}n's theorem, ][]{alfven}.
The Ohmic dissipation constraint is a sensible limit to adopt for regions with $R_\textrm{m} \lesssim 1-10$, where the behaviour of the model is somewhat predictable. This assumption is helpful to support the negligible evolution of the magnetic field in shallow regions when considering the induced Ohmic dissipation, thereby neglecting the non-linearity between the two mechanisms; induction and dissipation. Of course, in practice, the feedback starts playing a big role with depth, and the structure of the magnetic field becomes more complex, significantly differing from the measurable external magnetic field.

In this paper we apply the Ohmic dissipation prescription to Uranus and Neptune in order to constrain the maximum penetration depth of the zonal winds on these planets. It is a similar approach to that of \citet{liu} and \citet{wicht}, who have constrained the wind depths on the gas giants using the planetary luminosity as a limit (the former), and the heat and  entropy flux budgets (the latter).
Although the constraints placed on the winds of  Jupiter and Saturn by \citet{liu} do not consider the fact that the total Ohmic dissipation could indeed surpass the surface luminosity $L_{\textrm{surf.}}$ by a factor $\lesssim$100 \cite[e.g. see][]{hewitt, wicht}, their results are consistent with the gravitational constraints placed on the maximum penetration depth by \citet{kaspi2018}. 
This is not surprising, considering that the total Ohmic dissipation is  proportional to the electrical conductivity of the flow materials, which in the case of Jupiter and Saturn, is mostly due to semi-conducting hydrogen. Since the electrical conductivity of semi-conducting hydrogen obeys an exponential law scaling with density and temperature (and therefore with depth), a power output of  $\sim\!\!100 \, L_{\textrm{surf.}}$ is easily surmountable over short distances.

Three key differences are evident when applying this method for the ice giants. The \textit{first} is that the planetary magnetic fields of Uranus and Neptune are non-axisymmetric with respect to the rotation axis and their measurements admit multipolar solutions, with the quadrupole and the octopole components as strong as the dipole \citep{holme}. 
This is definitely not the case for the gas giants. The dipole component is dominant in both gas giants, with a 10$^\circ$ axis tilt in Jupiter and almost a perfect alignment of the dipole axis and the rotation axis in Saturn, compared to a 60$^\circ$ and 47$^\circ$ tilt in Uranus' and Neptune's axes, respectively \citep{uranus_mag, neptune_mag}. Thus, reducing the problem to a dipole would be inaccurate. However, this tilt of the poloidal component ensures that the magnetic field lines are not aligned with the rotation axis. This is important since alignment of the two would theoretically induce no Ohmic dissipation, as discussed by \citet{glatz}.

The \textit{second} difference is that the gravitational moments $J_i$ of Jupiter and Saturn have been accurately  measured by Juno and Cassini missions, up to $J_{10}$ for Jupiter \citep{J_j} 
(as well as odd harmonics) and Saturn \citep{J_s}. 
Since higher order harmonics help constrain the radial density distribution of the planets in the outer layers, these regions of interior structure models of the gas giants are better known than that of Uranus and Neptune. 
The only spacecraft that has measured the gravitational fields of the ice giants was the Voyager II in the late 1980's, where only $J_2$ and $J_4$ were inferred, with large error bars \citep{J_u, J_n}. Although these were then improved by ground-based observations \citep{jacob1,jacob2,jacob3}, they still remain the only $J_i$'s determined with confidence.
Therefore, there are considerable differences in the accuracy of the gravitational fields of both sets of giants, which lead to larger ambiguities in determining the planetary bulk compositions and internal structures. These in return, can lead to noticeable differences in the estimates for the maximum penetration depth of winds on ice giants.

The \textit{third} difference is that the compositions of Uranus and Neptune are unknown. Unlike Jupiter and Saturn, which are primarily composed of hydrogen and helium in the outer layers, Uranus and Neptune are thought to have a significant fraction of "liquid ices" (which are not liquids nor ices) in their composition, like water H$_2$O, ammonia NH$_3$ and methane CH$_4$ \citep{hub1, hubbard}.In Jupiter and Saturn, the electrical conductivity due to semi-conducting hydrogen is expected to reach 1 Sm$^{-1}$  at a pressure-level of $\sim 10^5$ bar  \citep{kaspi2019}.  This pressure-level  corresponds to deeper regions in Uranus and Neptune. The electrical conductivity can reach a value of $\sim$ 1 Sm$^{-1}$ already at much lower pressures than in the gas giants due to the existence of water, which has a significant ionic contribution. It should be noted, however, that it
is still unknown how much water is present in the whole planet \citep{helled}, especially in the outer layers  above the dynamo generation region. This uncertainty presents a  challenge when determining the electrical conductivity profiles of the ice giants in shallow layers.


This work addresses the aforementioned complications and provides a detailed prescription for calculating electrical conductivity profiles for ideally mixed H$_2$--H$_2$O mixtures under planetary conditions, which is then extended to include He. This method is then used to calculate the electrical conductivity profiles of various Uranus and Neptune interior structure models and subsequently the total induced Ohmic dissipation as a function of depth. The results are  compared with the heat flux and entropy flux budgets permitted by the interior structure models in order to deduce the maximum penetration depth of zonal winds, which are assumed to penetrate the planet along cylinders parallel to the its rotation axis.  

Our paper is structured as follows: In Section \ref{sec:ohm} we describe the methods for calculating the total Ohmic dissipation and present the zonal wind and magnetic field models we employ in our calculations. 
We describe our electrical conductivity equations for H$_2$--H$_2$O mixtures in Section \ref{sec:cond}. Then, Section \ref{sec:int} expands on the interior structure models and the equations of state that we adopt in our calculations. Our results for radial electrical conductivity profiles and the total induced Ohmic dissipation for both planets are given in Section \ref{sec:res}. We discuss our findings in Section \ref{sec:dis} and give our concluding remarks in Section \ref{sec:con}.


\section{Ohmic Dissipation}
\label{sec:ohm}
Uranus and Neptune are fast rotators: Voyager II measured solid-body rotation periods of $17.24$hr and $16.11$hr, respectively, although these periodicities might not represent their bulk rotation \cite{helled_shape}. While the exact rotation periods remain unknown, they are expected to be of the order of the Voyager values.  Therefore, the strong Coriolis force is expected to suppress the motion of the interior flow in directions that are not aligned with the rotation plane and limit the variations in velocity along the rotation axis as well. 
It is unknown how far this principle applies to the zonal winds; whether they can sustain their behaviour into deep-seated regions, or are truncated at some depth due to different mechanisms.
If the winds were to penetrate inside the planets undisturbed, the increase in their electrical conductivity, combined with the increase in magnetic field strength, would generate more Ohmic dissipation per volume with depth. Hence, the built up total Ohmic dissipation can be used as an upper limit for the maximum penetration depth of winds
considering the energy and entropy available to the system.

\subsection{Calculation of Total Ohmic Dissipation}
We start by calculating the induced electrical current due to the zonal winds. In the presence of a magnetic field $\mathbf{B}$, electrical current density $\mathbf{j}$ is given by Ohm's law:
\begin{equation}
\mathbf{j} = \sigma (\mathbf{E} + \mathbf{U} \times \mathbf{B}), 
\end{equation}
where $\sigma$ is the electrical conductivity; $\mathbf{E}$ and $\mathbf{U}$ are the electrical and the velocity fields, respectively.
The induced Ohmic dissipation per volume associated with this current density is $\mathbf{j}^2/\sigma$. The total induced Ohmic dissipation is then the volume integral of this term:
\begin{equation}
P_{\mathrm{tot}} = \int_V \dfrac{\mathbf{j}^2}{\sigma}dV. 
\label{eq:ohm}
\end{equation}
Following in the footsteps of \citet{liu_phd} and decomposing the magnetic and electrical field into poloidal and toroidal components, we can rewrite the current density as:
\begin{equation}
\mathbf{j} = \sigma \,(\mathbf{E} + \mathbf{U}_T \times \mathbf{B}_P + \mathbf{U}_T \times \mathbf{B}_T + \mathbf{U}_P \times \mathbf{B}_P + \mathbf{U}_P \times \mathbf{B}_T). 
\label{eq:j_big}
\end{equation}
Due to the strong coriolis force, the fluid motions are dominant in the toroidal direction, such that $|\mathbf{U}_P| \ll |\mathbf{U}_T|$.  Thus, we can safely say that
$|\mathbf{U}_P \times \mathbf{B}_P| \ll |\mathbf{U}_T \times \mathbf{B}_P|$. 
Furthermore, the magnitude of the toroidal magnetic field $|\mathbf{B}_T|$ due to winds is comparable to $\sim \!\! R_\textrm{m}|\mathbf{B}_P|$ \citep{cao2017}. This is motivated by the fact that the external planetary magnetic field can only consist of the poloidal component in this decomposition,  

 and the induced toroidal magnetic field in shallow regions are generated by the interaction between the zonal winds and the poloidal magnetic field.
We calculate the magnetic Reynolds number associated with the zonal flows as \citep{cao2017}:
\begin{equation}
R_\textrm{m}(r,\theta) = \frac{\langle U_\varphi(r,\theta)\rangle H_\eta}{\eta} = \frac{\langle U_\varphi (r,\theta)\rangle \sigma^2 \mu_0 }{-\partial_r \sigma},
\end{equation}
where $\langle U_\varphi(r,\theta) \rangle$ is the rms zonal velocity at $(r,\theta)$:
\begin{equation}
\langle U\rangle = \left(\frac{1}{2} \int_0^\theta U(r,\theta)^2 \sin{\theta} d\theta\right)^{1/2},
\end{equation}
and $H_\eta$ is the scale height of magnetic diffusivity:
\begin{equation}
H_\eta = \frac{\eta}{\partial_r \eta}.
\end{equation}
The generation of the toroidal field is prompted by the decreasing magnetic diffusivity. Thus, in the outer regions where $\eta$ is large (i.e. $R_\textrm{m} < 1$) its magnitude is less compared to the magnitude of its poloidal counterpart, such that $|\mathbf{U}_P \times \mathbf{B}_T| \ll |\mathbf{U}_T \times \mathbf{B}_P|$. 
Thus, the current density in Eq. (\ref{eq:j_big}) reduces to:
\begin{equation}
\mathbf{j} \approx \sigma (\mathbf{E} + \mathbf{U}_\varphi \times \mathbf{B}_P + \mathbf{U}_\varphi \times \mathbf{B}_T). 
\label{eq:j}
\end{equation}
Using the solenoidality of the current density ($\nabla \cdot \mathbf{j} = 0$), neglecting the contribution from radial currents ($j_r \approx 0$) and bounding integration constants associated with the electrical potential,
\citet{liu_phd} expresses the contributing terms to the current densities as:
\begin{subequations}
\begin{align}
&j_\theta \approx  \frac{\sigma(r)}{r}\left(\frac{\partial}{\partial \theta} \int_r^R (\mathbf{U}_\varphi \times \mathbf{B}_P)_r dr' + r 
(\mathbf{U}_\varphi \times \mathbf{B}_P)_\theta\right) 
\label{eq:jthe}\\
&j_\varphi \approx \frac{\sigma(r)}{r}\frac{\partial}{\partial \varphi} \int_r^R (\mathbf{U}_\varphi \times \mathbf{B}_P)_r dr' .
\label{eq:jphi}
\end{align}
\end{subequations}
A more detailed explanation of the steps between Eq. (\ref{eq:j}) and Eq. (\ref{eq:jthe}, \ref{eq:jphi}) is included in Appendix \ref{sec:j}.
The total Ohmic dissipation above a radius $r'$ is then determined by plugging the above terms into Eq. (\ref{eq:ohm}), where the integration in the radial direction is from $r'$ to the planetary radius $R$.

Note that, we have implicitly assumed that the electrical conductivity profile is spherically symmetric, which is meaningful, since the internal structure models we use are so as well. For clarity, the contributing terms to the induced Ohmic dissipation are explicitly given in Appendix \ref{sec:biggest_boy}.

The total induced Ohmic dissipation can be used as an upper limit to demonstrate that zonal winds cannot sustain their behaviour in the deeper layers of the planets.
How is this limit defined? \citet{liu} have argued that the total induced Ohmic dissipation cannot surpass the planetary luminosity. 
However, the total Ohmic dissipation can indeed surpass the surface luminosity \citep{backus75, hewitt}. Following \citet{wicht}, the total dissipation $P_{\mathrm{tot}}$ is bounded by the total dissipative heating $\mathcal{E}_Q$:
\begin{equation}
P_{\mathrm{tot}} \lesssim  \mathcal{E}_Q = \int\limits_{r_i}^{R} \frac{Q_{A}}{-T/ (\partial T/ \partial r)} r^2 dr, 
\label{eq:epsilon}
\end{equation}
where $Q_A$ is the advective contribution to the total heat flux.
This comes with the assumption that the adiabatic cooling is roughly the same as dissipative heating at each layer. The integration is over the whole convective volume. Assuming that advection is the major contributor to the heat flux, it is given by:
\begin{equation}
Q_A \approx Q_i + \int\limits^R_{r_i} dr \int\limits_S dS \;\tilde{\rho} \tilde{T} \: \dfrac{(L_{\textrm{surf.}} - Q_i)}{\int\limits_V  dV \, \tilde{\rho} \tilde{T} },
\end{equation}
where $Q_i$ is the heat flux through the core boundary and  tilde denotes hydrostaticity and adiabaticity. The integrand is the volumetric heat source given that the convection is always assumed to be adiabatic. \citet{wicht} neglected the contribution of the core $Q_i$ because it occupies 10\% of the radius in the Jupiter interior structure models they use. In the models that we adopt, the core occupies between 0 and $\sim 37\%$ of Uranus or Neptune radii depending on the model \citep{helled, nadine, vazan}. 
Although the assumption of neglecting $Q_i$ for large cores is far from perfect, it would still remain the secondary term in the calculation .\footnote{If the contribution from $Q_i$ were comparable to that of the convective region (i.e. doubling the heat flux limit), the maximum penetration depth for every model is set back $\sim 1\%$ in Neptune radii (see Figure \ref{fig:n_ohm}).}

As noted by \citet{wicht}, another constraint can be placed on the system, namely the entropy flux limit. This constraint does not require that the adiabatic cooling cancels out the dissipative heating at each radius. However, it provides a looser constraint on the maximum penetration depth of the zonal winds. \citet{hewitt} place the constraint at:
\begin{equation}
P_{\mathrm{tot}} \lesssim \mathcal{E}_S  = \frac{T(r)}{T_0}  L_{\mathrm{surf.}}, 
\label{eq:entropy}
\end{equation}
where $T(r)$ is the temperature at radius $r$ and $T_0$ the temperature at the boundary of the convective envelope (which we take as the surface temperature).

\subsection{Planetary Magnetic Field}
In the absence of currents outside the planet, the external magnetic field becomes irrotational $(\nabla \times \mathbf{B} = 0)$ and can be decomposed through a potential field: $\mathbf{B} = -\nabla \Phi$. Combined with the solenoidality condition $(\nabla \cdot \mathbf{B} = 0)$, we can represent the scalar potential $\Phi$ as a solution to the Poisson equation $(\Delta \Phi = 0)$:
\begin{equation}
\Phi = R \sum_{l=1}^{\infty} \bigg(\frac{R}{r}\bigg)^{l+1} \sum_{m=0}^l P_l^m(\cos \theta)  \big(g_l^m \cos (m\varphi) + h_l^m \sin(m\varphi) \big), 
\end{equation}
where $R$ is the planetary mean radius, $P_l^m$ are the Schmidt normalized associated Legendre polynomials 
and $g_l^m$, $h_l^m$ are the Gauss coefficients in units of nT \citep{holme}. We are using the Gauss coefficients from Table 1 in \citet{stanley}.


We use the observed external magnetic field as the poloidal field and extrapolate it inwards. This is supported by the fact that $R_\textrm{m}$ stays small in the outer part of the planets and we can treat the magnetic field lines to be relatively diffusive.

\subsection{Zonal Winds}
\label{subsec:wind}
\begin{figure}
    \centering
    \includegraphics[width = 0.95\columnwidth]{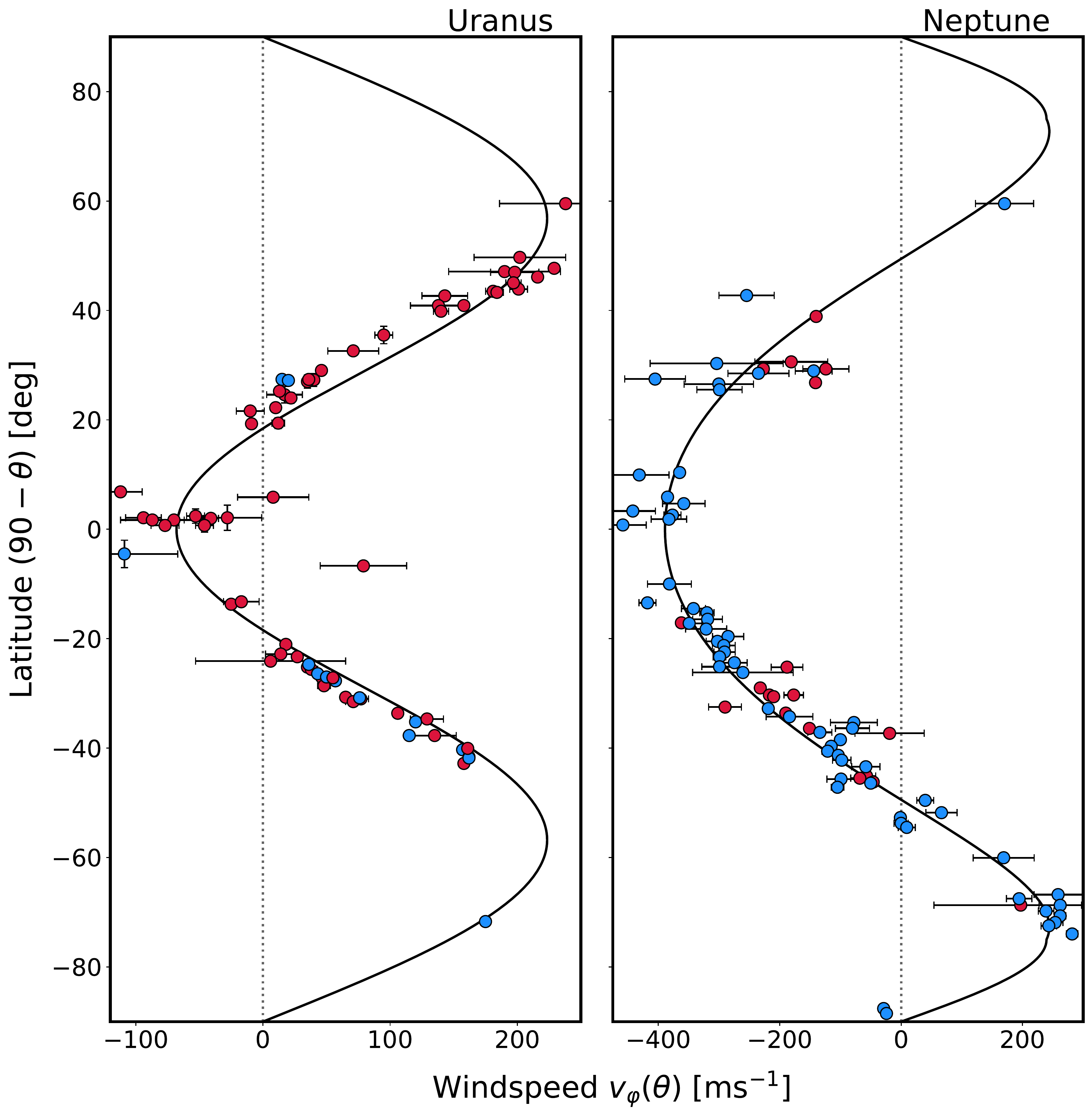}
     \caption{\textbf{Left Panel:} Zonal windspeeds in Uranus as a function of latitude. Red points are from Keck and Hubble Space telescope measurements \citep{hammel_keck, sromovsky} and blue points are Voyager II measurements \citep{hammel}. The solid line is the fit by the latter. \textbf{Right Panel:} Zonal winds speeds in Neptune as a function of latitude. Red points are from Hubble Space telescope measurements \citep{sro_nep} and blue points are Voyager II measurements \citep{lindal, limaye}. The solid line is the fit by \citet{french_wind}.}
     \label{fig:winds}
\end{figure}

Uranus and Neptune exhibit similar zonal wind profiles with a retrograde motion around the equator and prograde motion at higher latitudes as seen in Figure \ref{fig:winds}. Winds can reach up to $\sim 200$ ms$^{-1}$ on Uranus and $\sim 400$ ms$^{-1}$ on Neptune.
We use zonal wind profiles given by 
\citet{hammel} and
\citet{french_wind} for Uranus and Neptune, respectively. Both wind models are symmetric with respect to the equator and go to zero at the poles. 
Note that, small deviations in magnitude and direction have little to no contribution to the Ohmic dissipation in our calculations.
However, it is also important to remember that the rotation periods of the planets are given with respect to a solid-body rotation. Therefore, different rotation periods would result in different wind velocities \citep{helled_shape}. We address this later in Section \ref{sec:int}, when we introduce some interior structure models with differing rotation periods.

We assume that the zonal winds penetrate inside the planets on columns parallel to the rotation axis ($\boldsymbol{\omega} = \omega \,  \mathbf{e}_z$), where the azymuthal zonal wind velocity is not a function of z  ($\partial U_\varphi/\partial z = 0$), implying a Taylor-Proudman state for the flow  \citep{proudman, taylor}. Thus, the velocity $U_\varphi$ at a point $(r,\theta)$ in the planet (no-azymuthal dependence due to symmetry) is related to the observed surface zonal wind velocity by: 
\begin{equation}
U_\varphi (r, \theta) = v_\varphi \left(\arcsin\left({\frac{r \sin\theta}{R}}\right )\right).
\end{equation}

\section{Electrical Conductivity Profiles}
\label{sec:cond}
In giant planet interiors, the direct current (DC) electrical conduction mechanisms can be of two types.  
It can be either due to conducting electrons or due to charged ions being mobile and carrying an effective charge.
The former is the case when we are considering metals for instance, but it can also be the case of a semi-conductor being thermally excited where some electrons are being pumped into the conduction band.  
In a hydrogen--helium--water mixture, both H$_2$ and H$_2$O contribute to the total \textit{electronic} conductivity, however, the contribution of water turns out to be negligible compared to that of hydrogen. Thus, to first order we can write the total electrical conductivity as a sum of contributions of the semi-conducting hydrogen and ionically conducting hydrogen--water mixture:
\begin{equation}
\sigma_{\textrm{\scriptsize tot}} =  \sigma_{\textrm{\scriptsize{e$^-$}\!\!, \,\scriptsize{H$_2$}}} + \sigma_{\textrm{\scriptsize ion}} .
\end{equation}
Note, that this approximation neglects the effects of helium. The influence of helium is twofold:
\begin{enumerate}
    \item Since He does not contribute to the electrical conductivity of the mixture, the conductivity is reduced directly by its abundance.
    \item He atoms introduce more scatterers into the system, hence lowering the conductivity of the H$_2$--H$_2$O mixture even further.
\end{enumerate}
Below we develop estimates for the electronic and the ionic contributions to electrical conductivity and also describe the effects of helium for each contribution.

\subsection{Electronic Contribution}
\label{subsec:electronic}
Ideally, the electronic contribution to the conductivity should be provided by \textit{ab initio} simulations. However, such simulations are costly and are very difficult to perform at low density. 
Instead, starting from a modified Drude model \citep{celliers}, one can show (see Appendix \ref{sec:electronic} for details) that the electrical conductivity in the DC limit becomes:
\begin{equation}
    \sigma(0) = \sigma_0 e^{-E_\textrm{\scriptsize g}/2k_\textrm{\scriptsize B}T}, 
    \label{eq:elec_cond}
\end{equation}
where $E_\textrm{\scriptsize g}$ is the semi-conductor energy gap, $k_\textrm{\scriptsize B}$ the Boltzmann constant, $T$ the temperature, and $\sigma_0$ a prefactor. We stress here that both $\sigma_0$ and $E_\textrm{\scriptsize g}$ are functions of the density and the temperature.

The band gap in cryogenic hydrogen ($\rho$ = 0.0727 gcm$^{-3}$) is estimated at 15~eV. At higher density, we can use the experimental results by  \citet{nellis} who measured the conductivity of compressed hydrogen by gas gun shocks. For a typical density of $\rho = 0.27$ gcm$^{-3}$ they obtain a gap of $E_{\textrm{\scriptsize g}} =$ 11.7~eV and $\sigma_0=1.1\times10^8$~Sm$^{-1}$. This is a very high value for $\sigma_0$, four orders of magnitude higher than that expected from theory. 
Putting the results together and assuming a linear dependence of density for the band gap -- relatively acceptable assumption -- we get that the gap is given by:
\begin{equation}
    E_\textrm{\small g}(\rho)=16.216\;\textrm{eV} - 16.726\;\textrm{eV}\times \left(\frac{\rho}{\mathrm{1 gcm^{-3}}}\right).
    \label{eq:eg}
\end{equation}
If the modified Drude model is verified we should include a temperature and a density dependence in $\sigma_0$:
\begin{equation}
    \sigma_{0, \, \textrm{\scriptsize H$_2$}}=  1.1\times 10^8 \; \mathrm{Sm^{-1}} \left(\frac{\rho_\textrm{\scriptsize H$_2$}}{0.27 \mathrm{gcm^{-3}}}\right)^{-1/3} \left(\frac{T}{4000 \textrm{K}}\right). 
\end{equation}

The gap for water $E_{\textrm {\scriptsize g}, \, \textrm{\tiny H$_2$O}}$ is smaller than that of hydrogen, but  $\sigma_{0,\textrm{\tiny H$_2$O}}$ has been estimated to be much lower so that the ionic conductivity of water dominates over its electronic contribution.

However, it is unclear how much the presence of helium and water would reduce the conductivity of semi-conducting hydrogen. Since we are interested in a pressure regime (< 40 GPa), below where the metallization of hydrogen occurs, the effect of helium might not be significant. Nevertheless, water might reduce the electronic contribution of hydrogen dramatically \footnote{In any case, we find that the contribution from the semi-conducting hydrogen above $\sim 0.90R_{\scriptscriptstyle{\mathrm{U,N}}}$ is negligible compared to that of the ionically conducting  hydrogen--water mixture, even without taking into account the diminishing factors affecting the former.}.

\subsection{Ionic Contribution}

The fluctuation-dissipation theorem allows to associate the electrical conductivity to the auto-correlation function of the macroscopic currents:
\begin{equation} \label{eq:sigdef}
    \sigma=\frac{1}{3Vk_\textrm{\scriptsize B}T}\int_0^{+\infty}\textrm{d}t \left\langle \mathbf{J}(t)\cdot \mathbf{J}(0)\right\rangle,
\end{equation}
with the electrical current defined as:
\begin{equation}
    \mathbf{J}(t)=\sum_\alpha \sum_{i_\alpha} q_{i_\alpha}(t) \; \mathbf{v}_{i_\alpha}(t),
\end{equation}
where $q_{i_\alpha}$ is the instantaneous charge of the ${i_\alpha}^\textrm{\tiny th}$ particle of type $\alpha$, and $\vec{v}_{i_\alpha}$ its velocity. With this definition, assuming that electrons and ions are independent, which is the case for the Born-Oppenheimer limit, we find that:
\begin{equation}
    \sigma=\sigma_\textrm{\scriptsize ion}+\sigma_\textrm{\scriptsize  e$^-$}.
\end{equation}
Assuming that all the ions are independent, we get:
\begin{equation}
    \sigma_{\textrm{\scriptsize ion}}=\frac{1}{3Vk_\textrm{\scriptsize B}T}\sum_\alpha \sum_{i_\alpha}\int_0^{+\infty}\textrm{d}t \left\langle q_{i_\alpha}(t) \; \mathbf{v}_{i_\alpha}(t)\cdot q_{i_\alpha}(0) \; \mathbf{v}_{i_\alpha}(0)\right\rangle.
\end{equation}
With the further (much more uncertain) assumption that the charge of each species is time independent, we retrieve an Einstein-like relationship: 
\begin{equation}
    \sigma_\textrm{\scriptsize ion}=\sum_\alpha q_\alpha^2\frac{n_\alpha}{k_\textrm{\scriptsize B}T}D_\alpha,
    \label{eq:sigD}
\end{equation}
where $n_\alpha$ is the number density of species $\alpha$ and $D_\alpha$ its diffusion coefficient. For water,  \citet{french} found a relatively good agreement between Eq.~(\ref{eq:sigdef}) and Eq.~(\ref{eq:sigD}).

In a dissociating H$_2$--H$_2$O mixture, we expect  the following species to exist: H$_2$, H$_2$O, H$^+$, HO$^-$ and O$^{2-}$. 
Note that, H$_3$O$^+$ is totally negligible under the conditions of interest \citep{francois_diff}.
We assume that the charge is $0$ for H$_2$ and H$_2$O, $-1$ for HO$^-$  and $-2$ for O$^{2-}$. For the latter, this is a strong assumption because the conditions of appearance of O$^{2-}$ somewhat coincide with the condition for the metallization of oxygen \citep{mattsson}. 
Lone hydrogen is a more challenging case since it can come from either the dissociation of H$_2$ or of H$_2$O. In the first case, we can assume that hydrogen stays screened and has a charge close to 0 under the conditions of interest here. However, when it comes from the dissociation of water, since the oxygen atom is very electro-negative it keeps the electrons and the hydrogen is charged +1. We can mimic this difference with an effective charge for lone hydrogen. We define $\xi_{\alpha, \,\beta}$ as the fraction of atoms of type $\alpha$ in a species $\beta$. Thus, we have the conservation rule:  
\begin{equation}
    \sum_\beta \xi_{\alpha, \,\beta} = 1, \;\;\; \forall \, \alpha.
    \label{eq:cons}
\end{equation}
Since there are only two types of atoms, H, and O, using Eq. (\ref{eq:cons}), we can rewrite Eq. (\ref{eq:sigD}) as:
\begin{equation}\label{eq:sigD_H2H2O}
    \sigma_\textrm{\scriptsize ion}=
    \sum_\beta q_{\scriptstyle\beta}^2\frac{\xi_{\textrm{\scriptsize H}, \, \scriptstyle \beta} \, n_\textrm{\scriptsize H}}{k_\textrm{\scriptsize B}T}D_\beta +
    \sum_\beta q_{\scriptstyle\beta}^2\frac{\xi_{\textrm{\scriptsize O}, \,\scriptstyle\beta} \, n_\textrm{\scriptsize O}}{k_\textrm{\scriptsize B}T}D_\beta.
\end{equation}
However, the diffusion coefficient for each species is required to be able to compute this term. This is technically very intricate to obtain from \textit{ab initio} simulations. It is however possible to determine the diffusion of each atom type. 
The effective diffusion coefficient of each atom type $\alpha$ is given by:
\begin{equation}\label{eq:diffatom}
    \tilde{D}_\alpha= \sum_\beta \xi_{\alpha, \,\beta}D_\beta .
\end{equation}
Since oxygen is much heavier than hydrogen, we make the following assumption:
\begin{equation}
    \tilde{D}_{\textrm{\scriptsize O}}\simeq D_\textrm{\scriptsize H$_2$O}\simeq D_\textrm{\scriptsize HO$^-$}\simeq D_\textrm{\scriptsize O$^{2-}$}
    \label{eq:do}.
\end{equation}
For hydrogen, we assume a scaling by mass between H$^{\Theta+}$ (where $\Theta +$ is the effective charge of hydrogen) and H$_2$:
\begin{equation}
    D_\textrm{\scriptsize H$^{\Theta+}$}\simeq \sqrt{2} D_\textrm{\scriptsize H$_2$}.
\end{equation}
This is clearly not an optimal assumption, but it is expected to be of the same order as the previous ones. 
Using Eq.~(\ref{eq:diffatom}), we obtain:
\begin{align}
    \tilde{D}_\textrm{\scriptsize H} &= \xi_{\textrm{\tiny H},\textrm{\tiny H$^{\Theta+}$}}D_\textrm{\scriptsize H$^{\Theta+}$} + \xi_{\textrm{\tiny H},\textrm{\tiny H$_2$}}D_\textrm{\scriptsize H$_2$}+\xi_{\textrm{\tiny H},\textrm{\tiny HO$^-$}}D_\textrm{\scriptsize HO$^-$}+\xi_{\textrm{\tiny H},\textrm{\tiny H$_2$O}}D_\textrm{\scriptsize H$_2$O} \nonumber \\
    &\simeq \left( \xi_{\textrm{\tiny H},\textrm{\tiny H$^{\Theta+}$}} + \frac{\xi_{\textrm{\tiny H},\textrm{\tiny H$_2$}}}{\sqrt{2}}\right)D_\textrm{\scriptsize H$^{\Theta+}$}+ \left(\xi_{\textrm{\tiny H},\textrm{\tiny HO$^-$}}+\xi_{\textrm{\tiny H},\textrm{\tiny H$_2$O}}\right)\tilde{D}_{\textrm{\scriptsize O}}.
\end{align}
We can then extract the diffusion coefficient of the species H$^{\Theta+}$ as a function of quantities that can be calculated in simulations:
\begin{equation}
    D_\textrm{\scriptsize H$^{\Theta+}$}=\frac{1}{\xi_{\textrm{\tiny H},\textrm{\tiny H$^{\Theta+}$}}+\xi_{\textrm{\tiny H},\textrm{\tiny H$_2$}}/\sqrt{2}}\tilde{D}_{\textrm{\scriptsize H}}-\frac{\xi_{\textrm{\tiny H},\textrm{\tiny HO$^-$}}+\xi_{\textrm{\tiny H},\textrm{\tiny H$_2$O}}}{\xi_{\textrm{\tiny H},\textrm{\tiny H$^{\Theta+}$}}+\xi_{\textrm{\tiny H},\textrm{\tiny H$_2$}}/\sqrt{2}}\tilde{D}_{\textrm{\scriptsize O}}.
    \label{eq:dh}
\end{equation}
If we now use this last expression in Eq. (\ref{eq:sigD_H2H2O}), considering only the charged species, we get:
\begin{equation}
   \frac{k_\textrm{\tiny B}T}{e^2}\sigma_\textrm{\scriptsize ion} =  \Theta^2 \, n_{\textrm{\tiny H},\textrm{\tiny H$^{\Theta+}$}} \, D_\textrm{\scriptsize H$^{\Theta+}$} + n_{\textrm{\tiny H},\textrm{\tiny HO$^-$}}D_\textrm{\scriptsize HO$^-$} + 4 n_{\textrm{\tiny O},\textrm{\tiny O$^{2-}$}}D_\textrm{\scriptsize O$^{2-}$},
   \label{eq:simple_sig}
\end{equation}
with $\Theta$ the hydrogen effective charge. For the latter we assume that oxygen retains the hydrogen electron as water molecules dissociate because of the high electronegativity of oxygen. We also make the assumption that the resulting positive charge is distributed among all the lone hydrogen ions whether they come from dissociating water or hydrogen.  We thus have:
\begin{equation}
    \Theta=\frac{(2n_{\textrm{\tiny O},\textrm{\tiny O$^{2-}$}}+n_{\textrm{\tiny O},\textrm{\tiny HO$^-$}})}{ n_{\textrm{\tiny H},\textrm{\tiny H$^{\Theta+}$}}}. 
    \label{eq:pi}
\end{equation}
After introducing the molecular mixing ratio of water $z$ in a hydrogen--water mixture,
\begin{equation}
z  = \frac{n_\textrm{\tiny H$_2$O}}{n_\textrm{\tiny H$_2$}+n_\textrm{\tiny H$_2$O}} = \frac{2 n_\textrm{\tiny O}}{n_\textrm{\tiny H}}
\label{eq:z}
\end{equation}
 we can rewrite $\Theta$ as:
\begin{equation}
    \Theta=\frac{z}{2} \frac{2\xi_{\textrm{\tiny O},\textrm{\tiny O$^{2-}$}}+\xi_{\textrm{\tiny O},\textrm{\tiny HO$^-$}}}{ \xi_{\textrm{\tiny H},\textrm{\tiny H$^{\Theta+}$}}},
\end{equation}
in an ideal mixture. After some algebra, Eq. (\ref{eq:simple_sig}) can be expressed as:
\begin{align}
        \dfrac{k_\textrm{\tiny B}T}{e^2}\sigma_\textrm{\scriptsize ion} &= \dfrac{\xi_{\textrm{\tiny H},\textrm{\tiny H$^{\Theta+}$}}}{\xi_{\textrm{\tiny H},\textrm{\tiny H$^{\Theta+}$}}+\xi_{\textrm{\tiny H},\textrm{\tiny H$_2$}}/\sqrt{2}} \Theta^2 \left(n_{\textrm{\tiny H}}\tilde{D}_{\textrm{\scriptsize H}}-(\xi_{\textrm{\tiny O},\textrm{\tiny HO$^-$}}+2\xi_{\textrm{\tiny O},\textrm{\tiny H$_2$O}}) n_\textrm{\tiny O} \tilde{D}_{\textrm{\scriptsize O}}\right) \nonumber \\
        &+ n_\textrm{\tiny O}( \xi_{\textrm{\tiny O},\textrm{\tiny HO$^-$}} \tilde{D}_{\textrm{\scriptsize O}}+4\xi_{\textrm{\tiny O},\textrm{\tiny O$^{2-}$}}\tilde{D}_{\textrm{\scriptsize O}}).
        \label{eq:sig1}
\end{align}
Next we would like to include helium in the mixture and then compute the total ionic conductivity assuming  ideal mixing.
As mentioned above, helium has a twofold effect when it comes to decreasing the conductivity of the mixture. We do implement the first feature by adding helium to the mixture, thus reducing the abundance of hydrogen and water. However, we do not implement the secondary effect of introducing new scatterers to the system. 
Since the interplay between the temperature, molecular abundances, pressure and diffusion coefficients in a threefold mixture is not well understood, we choose to neglect the scattering effect of helium. 
Nevertheless, it is clear that helium would reduce the ionic conductivity. However, it is also important to note that since the conductivity of the mixture increases with the water content, using a fixed protosolar H$_2$:He ratio throughout the system, leads to smaller helium mass fractions when the conductivity is high.

We introduce the molecular mixing ratio of water in the hydrogen--helium--water mixture as
\begin{equation}
z_\textrm{\scriptsize tot} = \frac{n_\textrm{\tiny H$_2$O}}{n_\textrm{\tiny H$_2$}+n_\textrm{\tiny He}+n_\textrm{\tiny H$_2$O}},
\end{equation}
and analogously for hydrogen: $x_\textrm{\scriptsize tot}$ and helium: $y_\textrm{\scriptsize tot}$.
Finally, in the approximation of the ideal mixing law on the densities and ignoring the scattering effects of helium, we obtain the total ionic conductivity of the hydrogen--helium--water mixture:
\begin{align}
    \dfrac{k_\textrm{\tiny B}T}{e^2}\sigma_\textrm{\scriptsize ion} &= \Biggl[ \dfrac{\xi_{\textrm{\tiny H},\textrm{\tiny H$^{\Theta+}$}}}{\xi_{\textrm{\tiny H},\textrm{\tiny H$^{\Theta+}$}}+\xi_{\textrm{\tiny H},\textrm{\tiny H$_2$}}/\sqrt{2}} \Theta^2 \left(2\tilde{D}_{\textrm{\scriptsize H}}-(\xi_{\textrm{\tiny O},\textrm{\tiny HO$^-$}}+2\xi_{\textrm{\tiny O},\textrm{\tiny H$_2$O}}) z \tilde{D}_{\textrm{\scriptsize O}}\right) \nonumber \\
    &\hspace{1.3cm}+ ( \xi_{\textrm{\tiny O},\textrm{\tiny HO$^-$}} +4\xi_{\textrm{\tiny O},\textrm{\tiny O$^{2-}$}}) z \tilde{D}_{\textrm{\scriptsize O}} \Biggr] \; \frac{N_\textrm{A} (x_\textrm{\scriptsize tot} + z_\textrm{\scriptsize tot})}{V_\textrm{\scriptsize tot}},
    \label{eq:bigboy}
\end{align}
where $N_\textrm{A}$ is the Avogadro constant and $V_\textrm{\scriptsize tot}$ is the total ideally mixed molar volume;
\begin{equation}
V_\textrm{\scriptsize tot} = x_\textrm{\scriptsize tot} \frac{M_\textrm{\tiny H$_2$}}{\rho_\textrm{\tiny H$_2$}} + y_\textrm{\scriptsize tot} \frac{M_\textrm{\tiny He}}{\rho_\textrm{\tiny He}} + z_\textrm{\scriptsize tot} \frac{M_\textrm{\tiny H$_2$O}}{\rho_\textrm{\tiny H$_2$O}},
\end{equation}
where $M_i$ is the molar mass of species $i = $ H$_2$, He or H$_2$O and $\rho_i$ are the densities of the pure species, given by their respective equations of state. The densities and thus the total volume $V_\textrm{\scriptsize tot}$, the diffusion coefficients $\tilde{D}_\alpha$ and the different fractions $\xi_{\alpha,\,\beta}$ are all functions of pressure and temperature. The diffusion coefficients and the fractions also depend on the composition.

Since water dissociation is the most important contributor to the ionic conductivity and since it has been extensively studied by \citet{french2010}, we compared our conductivity predictions for pure water with those of their study. Using Eq.~(\ref{eq:bigboy}) with the fractions and diffusion coefficients of the following paragraphs, we obtain a good agreement in the $\sim1-50$ GPa pressure regime. For instance, at 2gcm$^{-3}$ and 6000 K, we obtain a conductivity of $\sim 13000$ Sm$^{-1}$, compatible with the slightly lower 9000 Sm$^{-1}$ obtained by \citet{french2010}. Experimental values are also similar to our calculations, where \citet{nellis} reports 2500 Sm$^{-1}$ at 40 GPa and 2600 K which is close to the $\sim3300$ Sm$^{-1}$ that we compute under similar conditions. The model used here provides reliable estimates of the water conductivity under the conditions of interest for the Ohmic dissipation in Uranus and Neptune.

\begin{figure*}
    \centering
    \includegraphics[width = 1.82\columnwidth]{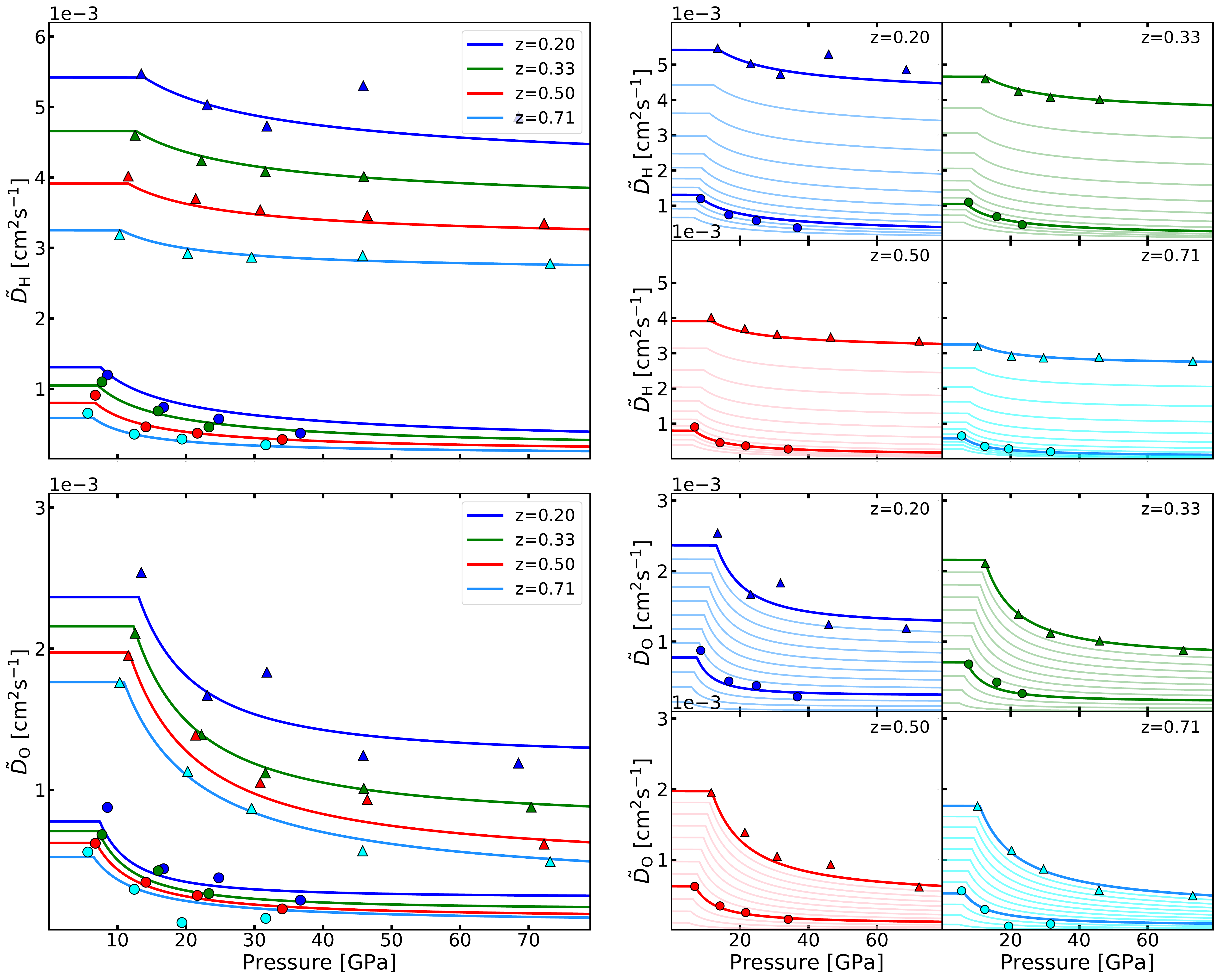}
    \caption{\textbf{Left Panels:} Fits to the hydrogen (top) and oxygen (bottom) diffusion coefficients from \citet{francois_diff} in two temperature regimes $T=2000$K (circles) and $T=6000$K (triangles) for different molecular mixing ratios $z= n_{\mathrm{H_2O}}/(n_\mathrm{H_2O}+n_\mathrm{H_2}) = \{0.20, 0.33, 0.50, 0.71\}$ using Eq. (\ref{eq:diff_h}) for hydrogen and  Eq. (\ref{eq:diff_o}) for oxygen (denoted as $x$ in that paper). \textbf{Right Panels:} Temperature scaling of the diffusion coefficients for constant molecular mixing ratios. The fainter lines represent different temperatures regimes with an increment of $\Delta T = 500$K between lines. The bold lines are the same fits shown on the left panels for temperatures $T = 2000$K and $T = 6000$K.}
    \label{fig:diff_h}
\end{figure*}

\subsubsection{Number Fractions $\xi_{\alpha, \, \beta}$}
The fractions of atoms $\alpha$ in various species $\beta$: $\xi_{\alpha, \, \beta}$ in  Eq.~(\ref{eq:bigboy}) are estimated by first approximating the dissociation fractions $\epsilon$ for the three dissociation reactions H$_2$~$\leftrightarrow$~2~H, H$_2$O~$\leftrightarrow$~OH$^{-}$+~H$^{+}$ and OH$^{-}\leftrightarrow$~O$^{2-}$+~H$^{+}$  using a hyperbolic tangent function:
\begin{equation}
\epsilon_{\alpha} = \frac{1}{2} \left(\tanh{\left(\frac{P_\alpha-P}{ \delta P_\alpha}\right)} + 1\right),
\label{eq:diss}
\end{equation}
where we model the pressures $P_\alpha$ and $\delta P_\alpha$ of different molecules $\alpha$ as functions of temperature (see Appendix \ref{sec:frac}). After calculating the dissociation fractions, the number fractions of H and O in different species can be calculated depending on the molecular mixing ratios $z$.  Figure \ref{fig:frac} shows the fraction of H in species H$^+$, H$_2$, H$_2$O, OH$^-$ and fraction of O in species H$_2$O, OH$^-$ and O$^{2-}$ in a H$_2$--H$_2$O mixture under various temperatures and molecular mixing ratios as a function of pressure using Eq.~(\ref{eq:diss}). We use the results of  \citet{francois_diss} for these ratios as a reference, and find that Eq.~(\ref{eq:diss}) is rather reliable for estimating the dissociation fractions of the molecules. The number fractions are especially in good agreement with the data in the low pressure regime ($P < 20$ GPa), which is important, since pressures above 20 GPa are reached around $\sim$0.80R$_{\scriptscriptstyle{\mathrm{U}}}$ and $\sim$0.85R$_{\scriptscriptstyle{\mathrm{N}}}$.

\subsubsection{Diffusion Coefficients $\tilde{D}_{\mathrm {\tiny{{H,O}}}}$}

Looking at Eq.~(\ref{eq:bigboy}), we see that the effective diffusion coefficients $\tilde{D}_\textrm{\scriptsize H}$ and $\tilde{D}_\textrm{\scriptsize O}$ are required in order to calculate the ionic conductivity of the mixture. As explained above, these could be approximated by the measurable quantities given in Eqs.~(\ref{eq:do}) and (\ref{eq:dh}).

\citet{francois_diff} have calculated the diffusion coefficients of hydrogen and oxygen in a H$_2$--H$_2$O mixture as a function of pressure using \textit{ab initio} simulations. Two different temperatures were considered: $T = 2000$K and $T = 6000$K, as well as four different molecular mixing ratios $z  = n_{\mathrm{H_2O}}/(n_\mathrm{H_2O}+n_\mathrm{H_2})$ = $\{$0.20, 0.33, 0.50, 0.71$\}$. The measured diffusion coefficients in the simulations are shown on the left panels of Figure \ref{fig:diff_h}.

In order to generalize the diffusion coefficients in temperature  and molecular mixing ratio $z$, we first fit a single power-law in pressure $P$ for constant $T$ and $z$. Then we choose the best fit as our anchor point and generalize the fit in $T$ and $z$ by modifying the single power-law.


Similarly to the number fractions, we can fit the \textit{ab initio}-simulation-derived diffusion coefficients as a function of the pressure $P$, temperature $T$ and molecular mixing ratio $z$. For both species, the diffusion coefficient for given temperature and composition is well described by a single power-law as a function of the pressure. Including, $z$ and $T$ to the fit we found that the following prescriptions satisfactorily reproduced the \textit{ab initio} data:

\begin{align}
&\tilde{D}_\mathrm{H}(P,z,T) = (P_{0,\mathrm{H}}  g^3(T)+ b_\mathrm{H}  P^{\, a_\mathrm{H} f(z)})   \frac{\sqrt{g(T)}}{f(z)},
\label{eq:diff_h} \\[0.5em]
&\tilde{D}_\mathrm{O}(P,z,T) = (P_{0,\mathrm{O}}  + b_\mathrm{O}  P^{\, a_\mathrm{O} h(z)})  h^{3}(z)  g^{3/2}(T),
\label{eq:diff_o}
\end{align}
where the modifying functions are given by:
\begin{equation}
f(z) = \alpha + \beta z, \hspace{0.5cm}h(z) = (\gamma + \delta z)^{-1}, \hspace{0.5cm}g(T) = \frac{T}{6000\mathrm{K}}.\label{eq:diff_mod}
\end{equation}
The different coefficients of the fit are provided in Appendix \ref{sec:coeff} for reproducibility. Despite its simplicity, this method nevertheless allows us to scale the diffusion coefficients.

Lastly, instead of extrapolating the diffusion coefficients to lower pressures than the regime considered in the \textit{ab initio} simulations, we take a conservative approach, and set their values as constants equal to the value calculated at the lowest pressure point in the simulation. Extrapolating the curves beyond the measurement points would yield higher values of electrical conductivity in lower pressures, but naturally introduces an uncertainty in the results. Therefore, in order to keep our energy/entropy budget violation depths as upper bounds, we adopt the conservative approach of cutting the curves. We find that in both cases the inferred values are similar around $\sim 0.9 R_{\mathrm{\scriptscriptstyle{U,N}}}$ for both planets.

Figure \ref{fig:diff_h} shows the behavior of the diffusion coefficients as a function of $P$ for different temperatures and molecular ratios. The agreement between the \textit{ab initio} data and the resulting fit is globally very good. The very simple form of the fit allows us to then easily determine the electrical conductivity for various conditions.

\section{Interior Models}
\label{sec:int} 
The electrical conductivity calculations strongly depend on the interior structure models and the used equations of state (EOSs). Since the internal structures of Uranus and Neptune are not well constrained, and therefore their compositions and thermal profiles can significantly vary, it is necessary to consider different models and interpret the results accordingly. 
Moreover, it is crucial to adopt reliable and up-to-date EOSs since they can affect the inferred conductivities. Below, we describe the interior structure models and EOSs we use for our  calculations.

\subsection{Equations of State (EOS)}
\label{subsec:eos}
We adopt the EOS developed by \citet{cms} for hydrogen and helium. For hydrogen and for the temperature regime that we are interested in ($T < 10^5$ K), it combines the \citet{scvh} EOS (i.e. SCvH EOS) for densities $\rho \le 0.05$ gcm$^{-3}$ with that of the \citet{eos2} for $0.3 < \rho \le 5.0$ gcm$^{-3}$, the latter based on \textit{ab initio} simulations. The gaps in  density between these two sets are then interpolated via a bicubic spline to maintain continuity of the function up to its second derivative. 
Similarly for helium two EOSs are combined for our temperature regime ($T < 10^6$ K), the SCvH EOS for densities $\rho \le 0.1$ gcm$^{-3}$ and the EOS by F.~Soubiran et al.~(2020, in preparation) for $1.0 < \rho \le 100.0$ gcm$^{-3}$. Again, a bicubic interpolation is used to smoothly stitch the EOS together.

For water, we adopt the EOS by \citet{shah} (submitted to A\&A). They combine various EOSs of water from \citet{water1} and \citep{water2} for $T < 200$ K and switch to \citet{mazevet} EOS at pressure $P = 1$ GPa for higher temperatures. The latter is based on \textit{ab initio} simulations.

The different EOSs are combined using an isothermal-isobaric ideal volume law. Under the range of conditions explored in the current work it has been shown to be a very good approximation, for hydrogen--helium \citep{cms}, for water--hydrogen \citep{francois_diss} and for ternary mixtures \citep{francois_HHeZ}.
\begin{figure}
    \centering
    \includegraphics[width=1\columnwidth]{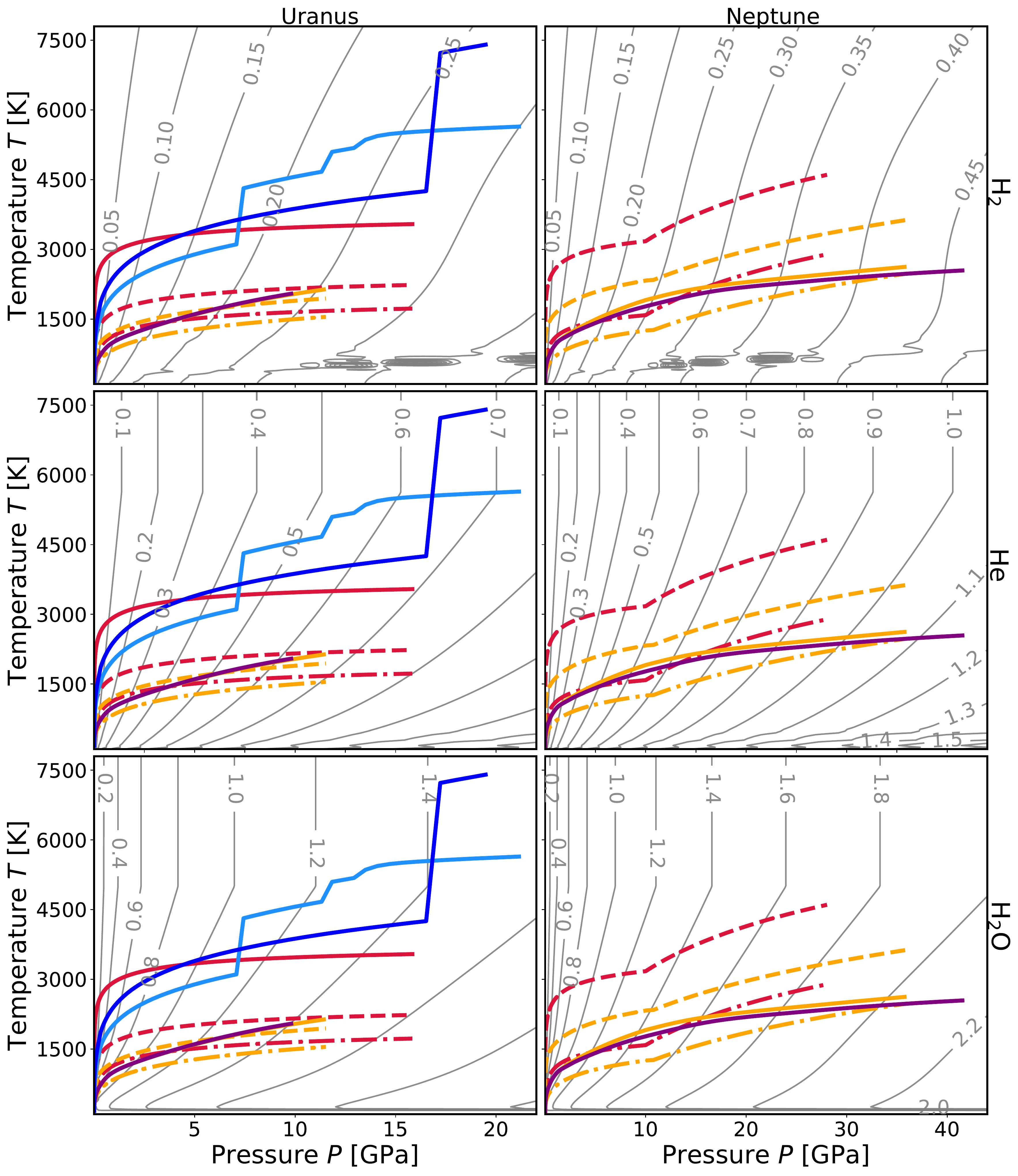}
    \caption{Equations of state of hydrogen (top), helium (center) and water (bottom) as a function of pressure and temperature. The contours represent the densities $\rho_i(P,T)$ of each species and the various colored curves show the interior structure models that we adopt in our study (down to $0.8 R_{\mathrm{\scriptscriptstyle{U,N}}}$). The contour lines have been kept to minimum and the legend is excluded for clarity. The color schemes and the linestyles correspond to those in Figures \ref{fig:u_rpt} for Uranus and Figure \ref{fig:n_rpt} for Neptune.}
    \label{fig:EOS}
\end{figure}
For illustrative purposes, Figure \ref{fig:EOS} shows where the interior structure models lie on the $(P,T)$ plane of the hydrogen, helium and water EOS, down to $0.8 R_{\mathrm{\scriptscriptstyle{U,N}}}$ for both planets. We elaborate on all the models we adopt in the following section.

\subsection{Interior Structure}
\subsubsection{Uranus}

\begin{figure}
    \centering
    \includegraphics[width=1\columnwidth]{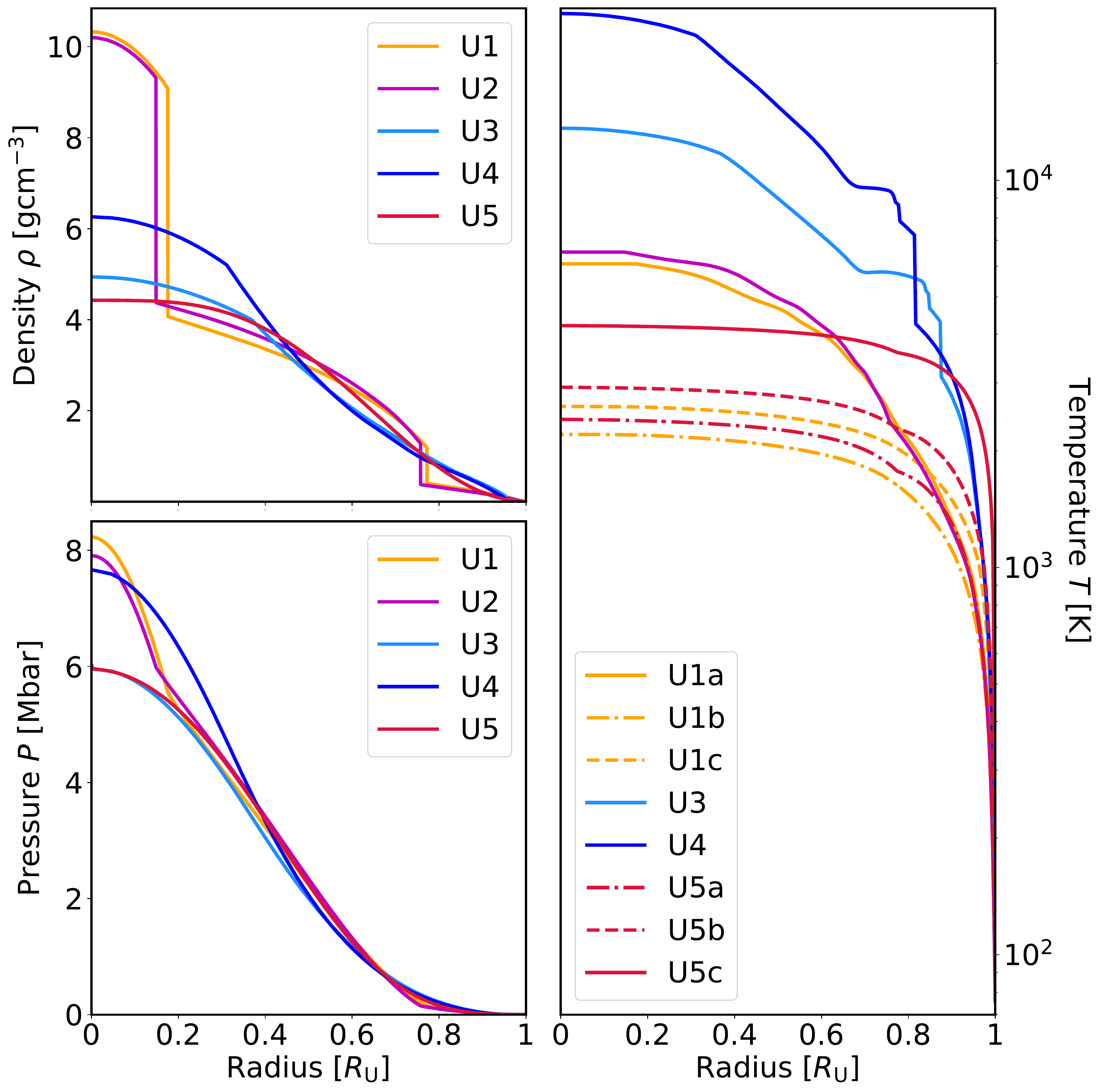}
    \caption{Interior structure models of Uranus. \textbf{Left Panels:} The density and the pressure profiles taken from  \citet{nadine}: U1, U2, \citet{vazan}: U3, U4 and \citet{helled}: U5. \textbf{Right Panel:} The temperature profiles corresponding to the density/pressure profiles. We use models with the original adiabatic temperature profiles from \citet{nadine} (solid orange for U1a, solid purple for U2), modified versions of U1 by \citet{podolak} (orange dot-dashed for single layer convection: U1b, dashed for 10$^6$ layer convection: U1c), the original temperature profiles from \citet{vazan} (light blue for U3 and dark blue for U4), and lastly  the assigned temperature profiles to the empirical model U5 by \citet{podolak} (red dot-dashed for single layer convection: U5a, dashed for 10$^6$ layer convection: U5b, solid for 10$^7$ layer convection: U5c).}
    \label{fig:u_rpt}
\end{figure}

For Uranus, we investigate five different density/pressure profiles with a total of nine associated temperature profiles, summarized in Table \ref{tab:u_tab}. The first set of density/pressure profiles, U1 and U2, are the ones presented by \citet{nadine}. These are adiabatic three-layer structure models, with a rocky core, an inner water envelope and a hydrogen-helium atmosphere with heavy elements. 
Both U1 and U2 are very similar with the main difference being the rotation period, where the former has a Voyager II measured period and the latter has a modified rotation period, inferred from the minimization of the dynamical heights of Uranus \citep{helled_shape}. The resulting modified rotation is $\sim 4 \%$ faster than the Voyager II measured period of 17.24 h.
Thus, the surface winds would be modified according to:

\begin{equation}
v_\varphi^{\scriptscriptstyle\textrm{new}}(\theta) = v_\varphi(\theta) + 2\pi \left(\frac{1}{T^{\scriptscriptstyle\textrm{new}}} - \frac{1}{T}\right) R_{\scriptscriptstyle\textrm{U,N}}  \sin{\theta},
\label{eq:wind_change}
\end{equation}
where $T$ is the rotation period and $\theta$ is the colatitude.\footnote{However, we find that the total Ohmic dissipation is actually not very sensitive to this correction. This also makes sense intuitively, since the change in the surface wind strength is just a small correction factor which would shift the total Ohmic dissipation profile by an insignificant amount.}
\begin{table*}
	\centering
	\caption{Overview of the interior structure models of Uranus. Model numbers correspond to different density vs. pressure profiles. Letters following the model number (e.g. U1a, U1b, U1c) indicate a different temperature profile is assigned to the same density/pressure profile. The original names of the models in their respective papers are listed on the rightmost column for clarity. }
	\begin{tabular}{l | c c c c} 
		\hline
		 & Density/Pressure Profile &  Temperature Profile & Convective Layers & Original Name \\
		\hline
		U1a & \citet{nadine} & \citet{nadine} & 1 & U1\\
		U1b & \citet{nadine} & \citet{podolak} & 1 & U1 Cold Model\\
		U1c & \citet{nadine} & \citet{podolak} & 10$^6$ & U1 Hot Model\\
		U2  & \citet{nadine} & \citet{nadine} & 1 &  U2\\
		U3  & \citet{vazan} & \citet{vazan} & -- &  V3\\
		U4  & \citet{vazan} & \citet{vazan} & -- &  V4\\
		U5a & \citet{helled} & \citet{podolak} & 1 &  PolyU Cold Model\\
		U5b & \citet{helled} & \citet{podolak} & 10$^6$ &  PolyU Hot Model\\
		U5c & \citet{helled} & \citet{podolak} & 10$^7$ &  --\\
		\hline
	\end{tabular}
	\label{tab:u_tab}
\end{table*}
In addition to the original U1, we use the original density-pressure profiles but when considering two different temperature profiles calculated by \citet{podolak} using the double diffusive convection prescription developed by \citet{LC}. To this purpose, \citet{podolak} assign a single layered and 10$^6$ layered convective model, shown by the dashed and the dot-dashed orange curves on the right panel of Figure \ref{fig:u_rpt}.

Next, we consider two originally non-adiabatic models, U3 and U4 by \citet{vazan}. These models are evolved with different primordial composition distributions and initial energy budgets, and then evolved to fit present day Uranus models. These models have been constructed to account for the possibility that Uranus' interior is still very hot and that its composition gradient is preventing the escape of heat effectively, explaining Uranus' measured  low-luminosity. The very high temperatures can be seen in Figure \ref{fig:u_rpt}, displayed by the blue curves. One key difference between the original U3 and U4 is their initial compositional difference, where the "metals" in the planet are made up of 2/3 H$_2$O + 1/3 SiO$_2$ in U3 and 1/3 H$_2$O + 2/3 SiO$_2$ in U4. 
\begin{table*}
	\centering
	\caption{Overview of the interior structure models of Neptune. Model numbers correspond to different density vs.~pressure profiles. Letters following the model number (e.g. N1a, N1b, N1c) indicate a different temperature profile is assigned to the same density/pressure profile. The original names of the models in their respective papers are listed on the rightmost column for clarity.}
	\begin{tabular}{l | c c c c} 
		\hline
		 & Density/Pressure Profile &  Temperature Profile & Convective Layers & Original Name \\
		\hline
		N1a & \citet{nadine} & \citet{nadine} & 1 & N1\\
		N1b & \citet{nadine} & \citet{podolak} & 1 & N1 Cold Model\\
		N1c & \citet{nadine} & \citet{podolak} & 10$^6$ & N1 Hot Model\\
		N2  & \citet{nadine} & \citet{nadine} & 1 &  N2b\\
		N3a & \citet{helled} & \citet{podolak} & 1 &  PolyN Cold Model\\
		N3b & \citet{helled} & \citet{podolak} & 10$^6$ &  PolyN Hot Model\\
		\hline
	\end{tabular}
	\label{tab:n_tab}
\end{table*}

Lastly, we investigate U5 by \citet{helled}, who represent the radial density distribution of the planet by a sixth-degree polynomial. It is originally developed as an "empirical" model providing only the pressure-density profile. This model fits the measured gravitational harmonics $J_2$, $J_4$ and bulk parameters like radius, mass and Voyager II measured solid-body rotation periods of Uranus but assumes no composition. This model is then assigned radial temperature profiles by \citet{podolak}, using again the double diffusive convection prescription analogous to the method for U1 mentioned above. To this purpose, \citet{podolak} assign a single, 10$^6$ and 10$^7$ layered convection models to U5, represented by the three red curves on the right panel of Figure \ref{fig:u_rpt}.


Next we derive the water content of the interior structure models  in shallow regions, using the EOS mentioned in the previous paragraphs. We assume that the outer regions of the planets are made up of hydrogen, helium, and water. We also assume ideal mixing for the system:

\begin{equation}
\frac{1}{\rho} = \frac{X}{\rho_\textrm{\tiny H$_2$}} + \frac{Y}{\rho_\textrm{\tiny He}} + \frac{Z}{\rho_\textrm{\tiny H$_2$O}},
\end{equation}
where $\rho$ is the radial density profile $\rho(r)$ making up the interior structure models. $X$, $Y$ and $Z$ are the mass fractions and $\rho_\textrm{\tiny H$_2$}$, $\rho_\textrm{\tiny He}$ and $\rho_\textrm{\tiny H$_2$O}$ are the density profiles of of hydrogen, helium, and water (determined for a given pressure and temperature: $\rho_i(P,T)$ from the EOS), respectively. We adopt a protosolar H$_2$:He ratio of $\chi^{-1} := X/Y = 0.745/0.255 $, by mass \citep{lod, podolak} throughout the region of interest. This assumption, combined with the fact that the mass fractions add up to $X+Y+Z = 1$, allows to compute a unique profile for $X(r)$ and $Z(r)$. The hydrogen mass fraction $X$ is then given by:

\begin{equation}
        X = \left(\frac{1}{\rho} - \frac{1}{\rho_\textrm{\tiny H$_2$O}} \right)\times\left(\frac{1}{\rho_\textrm{\tiny H$_2$}} 
        + \frac{\chi}{\rho_\textrm{\tiny He}} - \frac{1+\chi}{\rho_\textrm{\tiny H$_2$O}} \right)^{-1},
\end{equation}
where then $Z$ follows from $Z = 1 - X (1 + \chi)$.
By using the molecular mixing ratio of water defined in Eq.~(\ref{eq:z}) we can relate the hydrogen mass fraction to the molecular mixing ratio of water in a hydrogen--water mixture;

\begin{equation}
\frac{n_\textrm{\tiny H$_2$O}}{n_\textrm{\tiny H$_2$}} = \frac{z}{1-z} = \frac{Z}{X} \frac{M_\textrm{\tiny H$_2$}}{M_\textrm{\tiny H$_2$O}}.
\end{equation}
It is important to remember that the inferred compositions might  not be fully representative of the inner structure of Uranus, since we assume a constant H$_2$:He ratio and neglect any other species that are known to be present in outer regions of Uranus, like methane CH$_{4}$ and ammonia NH$_3$.

\subsubsection{Neptune}

For Neptune, we investigate three different density/pressure profiles with a total of six associated temperature profiles, summarized in Table \ref{tab:n_tab}. 

The first set of pressure-density profiles, N1 and N2, are by \citet{nadine}. The original temperature profiles of these models are (analogous to U1 and U2) adiabatic three-layer structure models, with a rocky core and two convective layers on top, consisting of hydrogen, helium and water. 
Again, the main difference being the rotation period, where the former has a Voyager II measured period and the latter has a modified rotation period, calculated via minimizing the dynamical heights of the winds on Neptune \citep{helled_shape}. The resulting modified rotation is $\sim 8 \%$ slower than the Voyager II measured period of 16.11 h. The surface winds are again corrected according to Eq.~(\ref{eq:wind_change}).

\begin{figure}
    \centering
    \includegraphics[width=1\columnwidth]{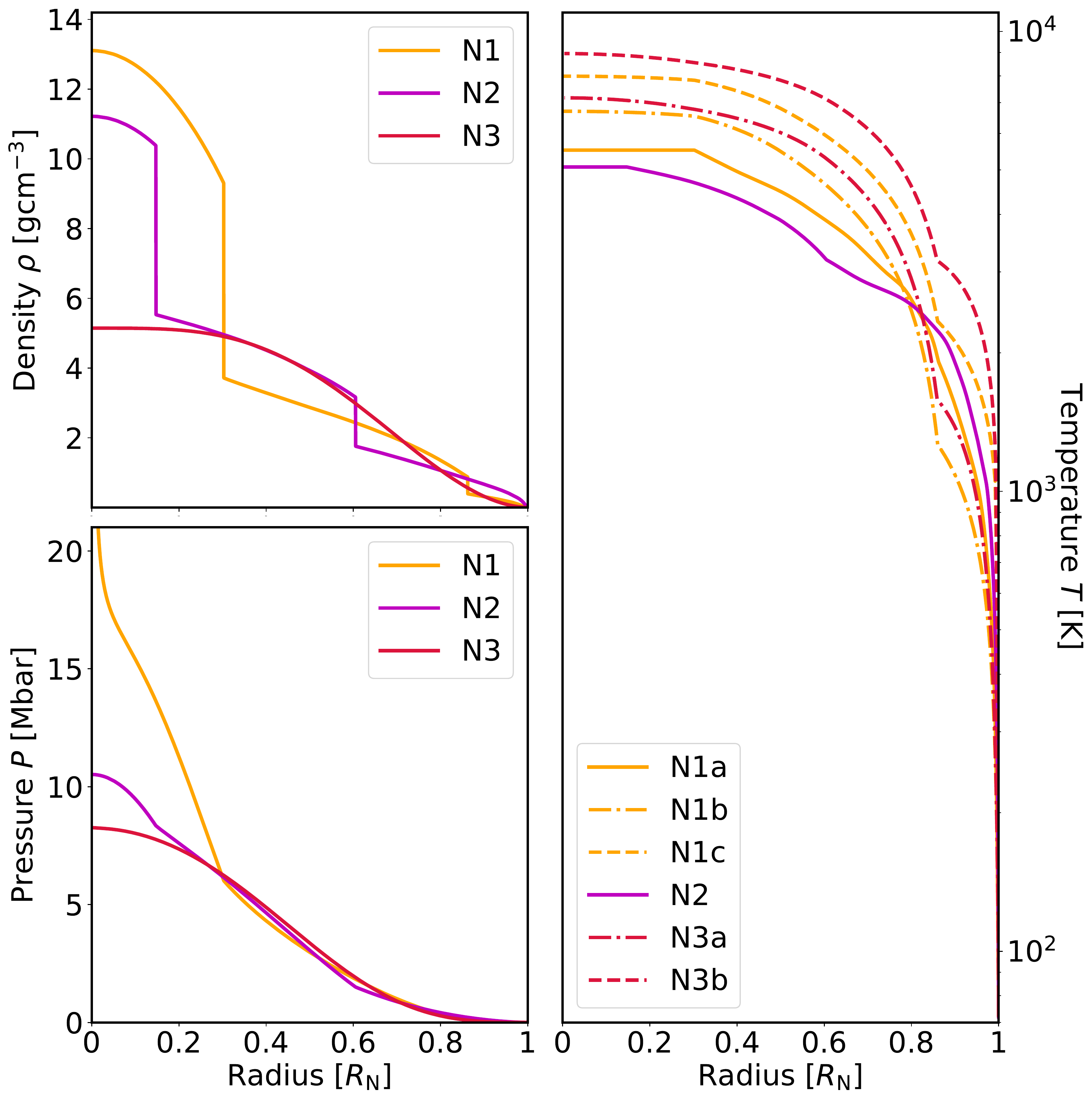}
    \caption{Interior structure models of Neptune. \textbf{Left Panels:} The density and the pressure profiles taken from  \citet{nadine}: N1, N2 and \citet{helled}: N3 \textbf{Right Panel:} The temperature profiles corresponding to the density/pressure profiles. We use models with the original adiabatic temperature profiles from \citet{nadine} (solid orange for N1a, solid purple for N2), modified versions of N1 by \citet{podolak} (orange dot-dashed for single layer convection: N1b, dashed for 10$^6$ layer convection: N1c), and the assigned temperature profiles to the empirical model N3 by \citet{podolak} (red dot-dashed for single layer convection: N3a and dashed for 10$^6$ layer convection: N3b)}
    \label{fig:n_rpt}
\end{figure}

In addition to N1 and N2, we consider two different temperature profiles by \citet{podolak} assigned to N1, analogous to the Uranus case. The modified temperature profiles represent single layered and 10$^6$ layered models, shown by the dashed and the dot-dashed orange curves on the right panel of Figure \ref{fig:n_rpt}. 

Lastly, we adopt N3 by \citet{helled}. This is again an empirical structure  model. 
These models are then assigned radial temperature profiles by \citet{podolak}, using the double diffusive convection prescription developed by \citet{LC}. To this purpose, \citet{podolak} assign a single and 10$^6$ layered convection models to N3, represented by the two red curves on the right panel of Figure \ref{fig:n_rpt}.

Using the same mixing prescription described in the previous paragraphs for Uranus, we calculate metallicity profiles $Z$ and the  water molecular mixing ratios $z$  associated with the interior structure models.

\section{Results}
\label{sec:res}
\subsection{Electrical Conductivity Profiles}

\begin{figure*}
    \centering
    \includegraphics[width = 1.73\columnwidth]{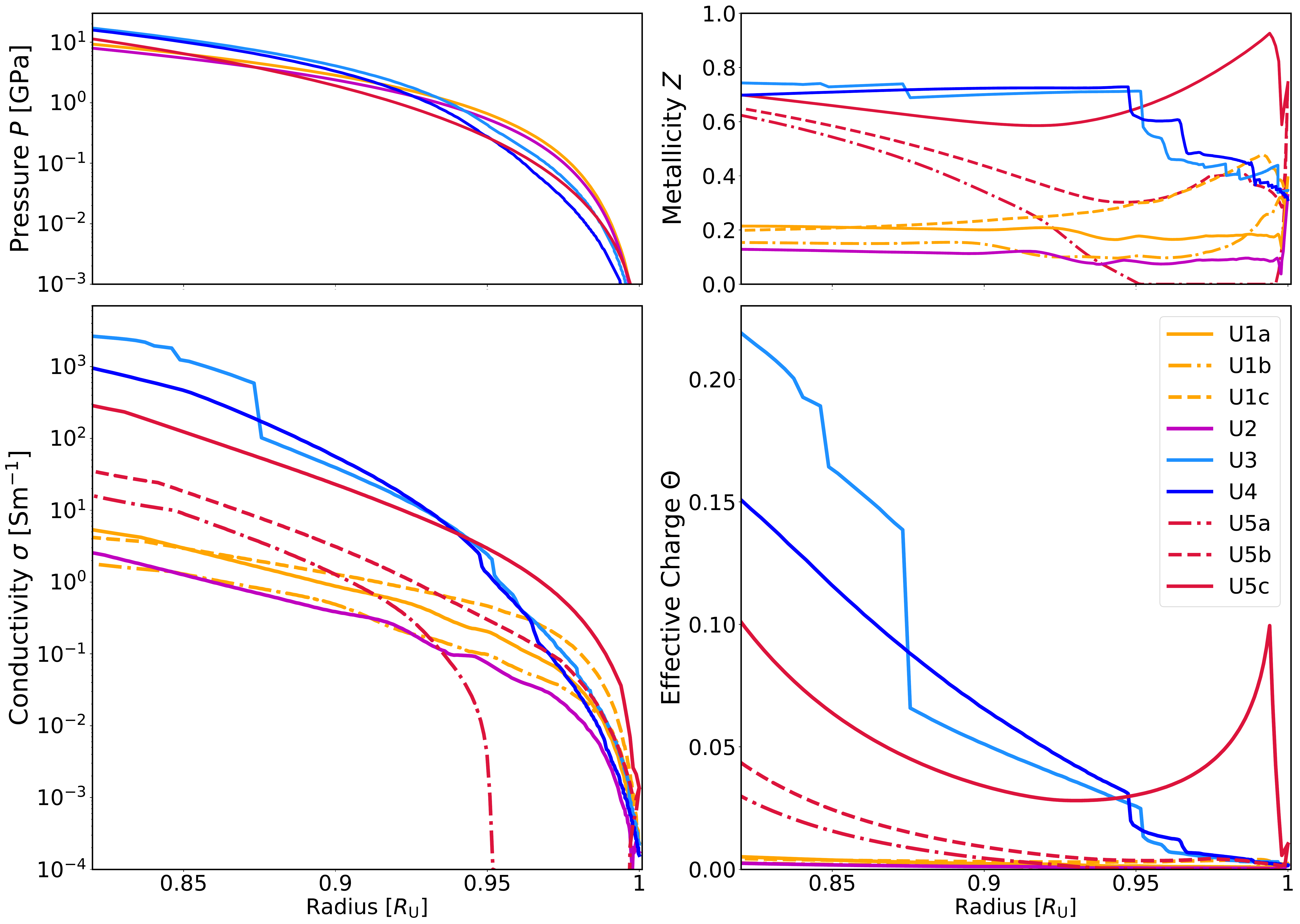}
    \caption{\textbf{Top Left Panel:} Pressure profiles of the Uranus structure models in the outer layers. \textbf{Top Right Panel:} Inferred metallicity  profiles of the Uranus models. \textbf{Bottom Left Panel:} Ionic conductivity profiles of the Uranus interior structure models.  \textbf{Bottom Right Panel:} Effective hydrogen charge $\Theta$ of the same models. It is clear that the models with high water abundance in the outer layers reach higher electrical conductivities at shallower layers. This is expected, since the effective charge $\Theta$ of the dissociated hydrogen becomes negligible as the water abundance decreases (see Eq. (\ref{eq:pi})). 
    }
    \label{fig:u_cond}
\end{figure*}
\begin{figure*}
    \centering
    \includegraphics[width = 1.73\columnwidth]{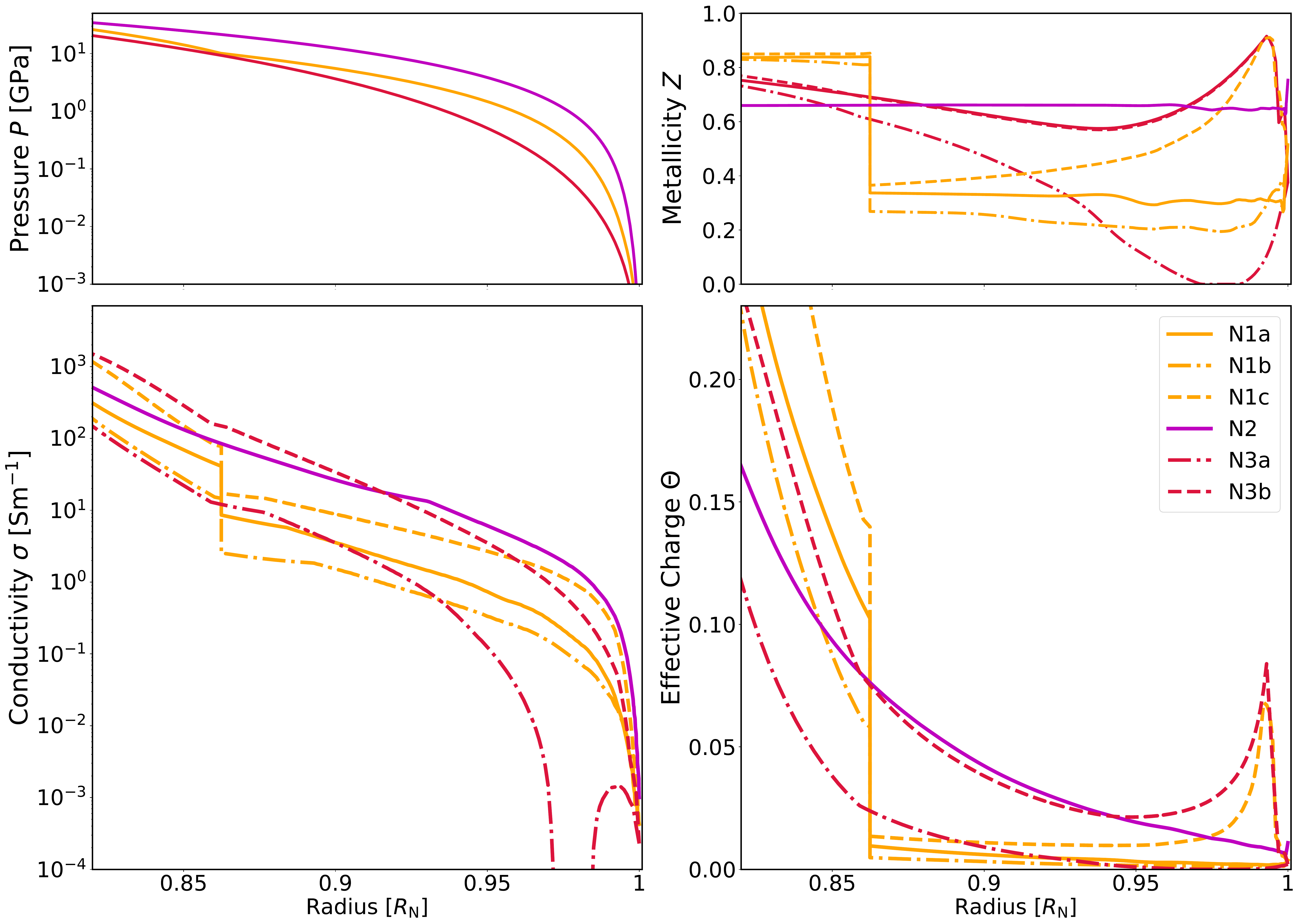}
    \caption{Same as Figure \ref{fig:u_cond}, but for Neptune.
    }
    \label{fig:n_cond}
\end{figure*}

The radial ionic conductivity profiles of various interior structure models of Uranus and Neptune are shown in Figure \ref{fig:u_cond} and \ref{fig:n_cond}, respectively, along with the inferred metallicity profiles. The electronic contribution of semi-conducting hydrogen is expected to be negligible compared to the ionic conductivity of the hydrogen--water mixture, due to the presence of other scatterers; helium and most importantly, water. 

A comparison between the conductivity plot and the metallicity profile shows that higher water content leads to higher electrical conductivity. This is expected, because both the effective charge of hydrogen $\Theta$ (see Eq.~(\ref{eq:pi})) and the contribution from the water ions (see Eq.~(\ref{eq:bigboy}) for OH$^-$ and O$^{2-}$) increases with increasing water content. The combination of higher pressures and temperatures in Neptune's shallow layers lead to higher electrical conductivity (for water-rich models) in the outermost layers in comparison to Uranus. 
However, it is not trivial to describe this trend quantitatively since the density, pressure and temperature profiles affect electrical conductivity prescription in a variety of ways, such as:
\begin{itemize}
\item the effective diffusion coefficients $\tilde{D}_\mathrm{H,O}$,
\item number fractions of atoms in different species $\xi_{\alpha, \beta}$,
\item the dissociation fraction of molecules $\epsilon_i$,
\item the bulk metallicities and their dependency on the EOS,
\item or the ionic conductivity equation itself (Eq. \ref{eq:bigboy}). 
\end{itemize}

This convoluted behaviour of the prescription then leads to our next point, which is that the electrical conductivity profiles are heavily model dependent. Looking at for example, around $0.9 R_{\mathrm{\scriptscriptstyle{U}}}$ in all the profiles (where we can argue that the atmospheric effects that are not considered become negligible), we find a difference of two orders of magnitude in conductivity between the various models.

\subsection{Ohmic Dissipation Profiles}
The total Ohmic dissipation associated with the aforementioned radial conductivity profiles are shown in Figures \ref{fig:u_ohm} and \ref{fig:n_ohm}, along with the planetary luminosity (pink-dashed), the energy flux limit (black-dashed), and the entropy flux limit (black-solid). 
Considering all the models, the overrun of the energy flux limit is found to be between $0.93 - 0.97 R_{\mathrm{\scriptscriptstyle{U}}}$ for Uranus, shallower than the depth where the conventional 3-layer models assume the transition to the ice-layer/envelope. For Neptune, the energy flux limit is surpassed between $0.95 - 0.98 R_{\mathrm{\scriptscriptstyle{N}}}$. The entropy flux limit, on the other hand, provides looser constraints on the maximum penetration depth of the winds, $0.90 - 0.95 R_{\mathrm{\scriptscriptstyle{U}}}$ for Uranus, and $0.92 - 0.97 R_{\mathrm{\scriptscriptstyle{N}}}$ for Neptune.

Despite the strong variance of electrical conductivity with different structure models, both Ohmic dissipation limits confine the maximum penetration depth estimates to relatively shallow regions in all of the models.
Interestingly, the four hottest structure models which have quite different electrical conductivity profiles: U1c, U3, U4 and U5c, all surpass the entropy flux limit around the \citetalias{kaspi} limit. This suggests that Uranus might be better described by hotter, non-adiabatic models with higher metallicities \citep{podolak, vazan}.

Taking a closer look at models which were constructed in similar fashion: U1a vs N1a, U1b vs N1b, U1c vs N1c, U5a vs N3a and U5b vs N3b, we see that Neptune models have higher electrical conductivity by an order of magnitude than Uranus'. 
Furthermore, for the considered pairs; the density/pressure profiles are modeled in the same way,
and the temperature profiles are modeled with the same prescription \citep{nadine, podolak}, with pairs having the same number of convective layers, but the overrun of Ohmic dissipation is $\sim 5\%$ shallower for the Neptune models, even though both the energy and entropy flux limits are an order of magnitude greater for Neptune than that of Uranus.

\section{Discussion}
\label{sec:dis}
The inferred Ohmic limits (and therefore the penetration depths) for structure models with the same density-pressure profile change with the assumed temperature profile.
This is expected, since both the limit itself (Eq.~(\ref{eq:epsilon}), (\ref{eq:entropy})) and the maximum penetration depth (via the conductivity Eq.~(\ref{eq:bigboy})) are temperature dependent. 
The study of \citet{kaspi} suggests that the maximum penetration depth of the winds is similar in both planets. Given that the two planets are not identical one could ask: What are the configurations in terms of temperature and composition  that would lead to a similar  Ohmic dissipation constraint at similar depths?

Our  results indicate that a similar penetration depth on both planets would be due to: (1) Higher temperatures in Uranus' shallow layers. This is consistent with various studies implying that Uranus' interior is non-adiabatic \citep{vazan, podolak, helled}, (2) Neptune's atmosphere is less water-rich, possibly resulting in a lower electrical conductivity.  This scenario is consistent with recent measurements of Neptune's atmospheric composition implying that this planet might be rock-dominated \citep{teanby}. 
Of course, it can be a combination of the two possibilities as well. 

\begin{figure*}
    \centering
    \includegraphics[width= 1.75\columnwidth]{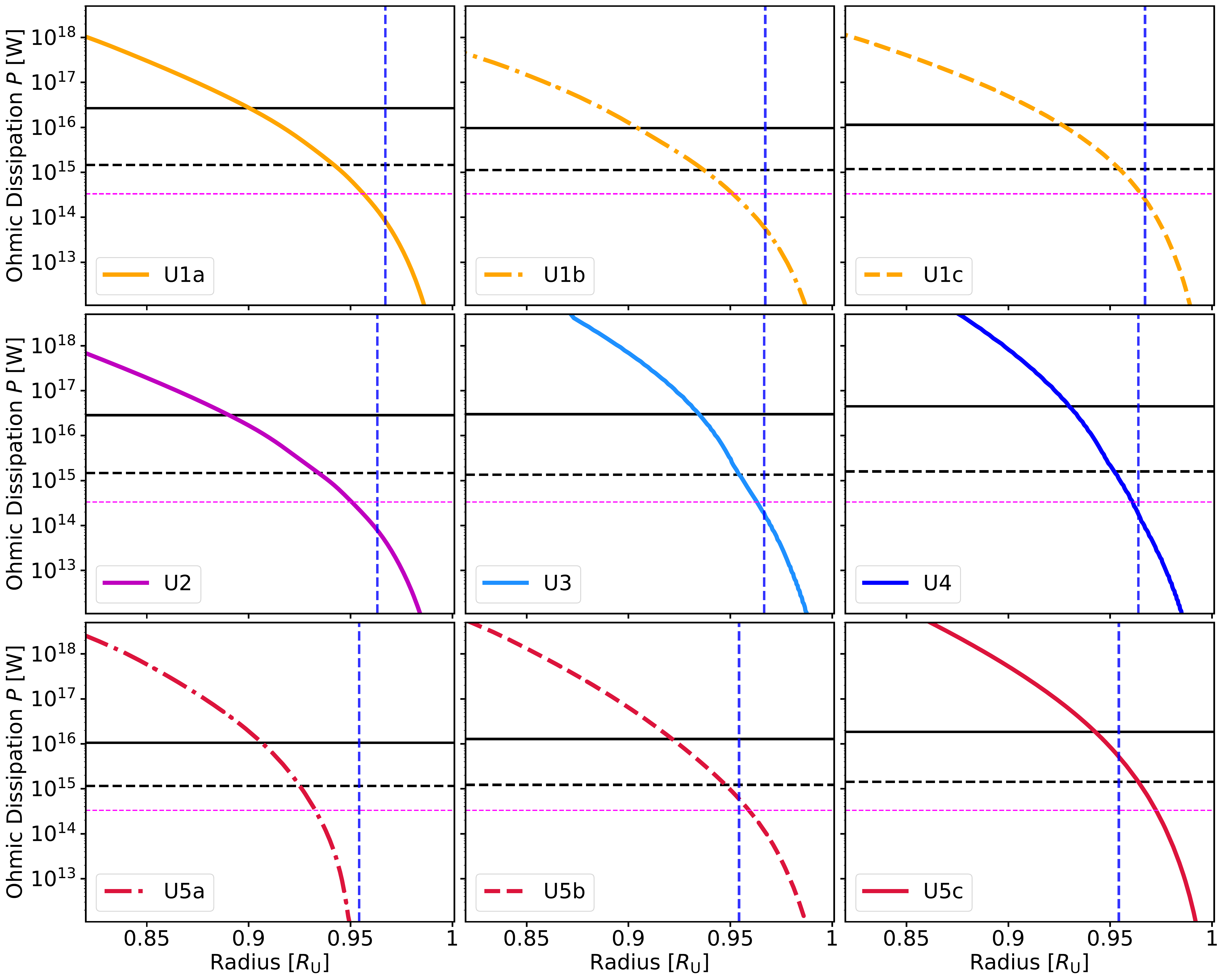}
    \caption{Inferred Ohmic dissipation profiles for Uranus. The horizontal dashed black line shows the energy flux limit given by Eq.~(\ref{eq:epsilon}) and the horizontal solid black line the entropy flux limit given by Eq.~(\ref{eq:entropy}). The planetary luminosity is also shown for illustrative purposes as the horizontal dashed-pink line. The vertical blue dashed lines represent the e-folding depth of the surface windspeeds, calculated using the method presented in \citetalias{kaspi} via the dynamical contribution of the winds to the gravitational harmonic $J_4$. We see that the energy flux constraint limits the wind penetration to $0.93 - 0.97 R_{\mathrm{\scriptscriptstyle{U}}}$, depending on the model. The entropy flux is less tight, constraining the penetration to depths between $0.90 - 0.95 R_{\mathrm{\scriptscriptstyle{U}}}$.}
    \label{fig:u_ohm}
\end{figure*}

\begin{figure*}
    \centering
    \includegraphics[width= 1.75\columnwidth]{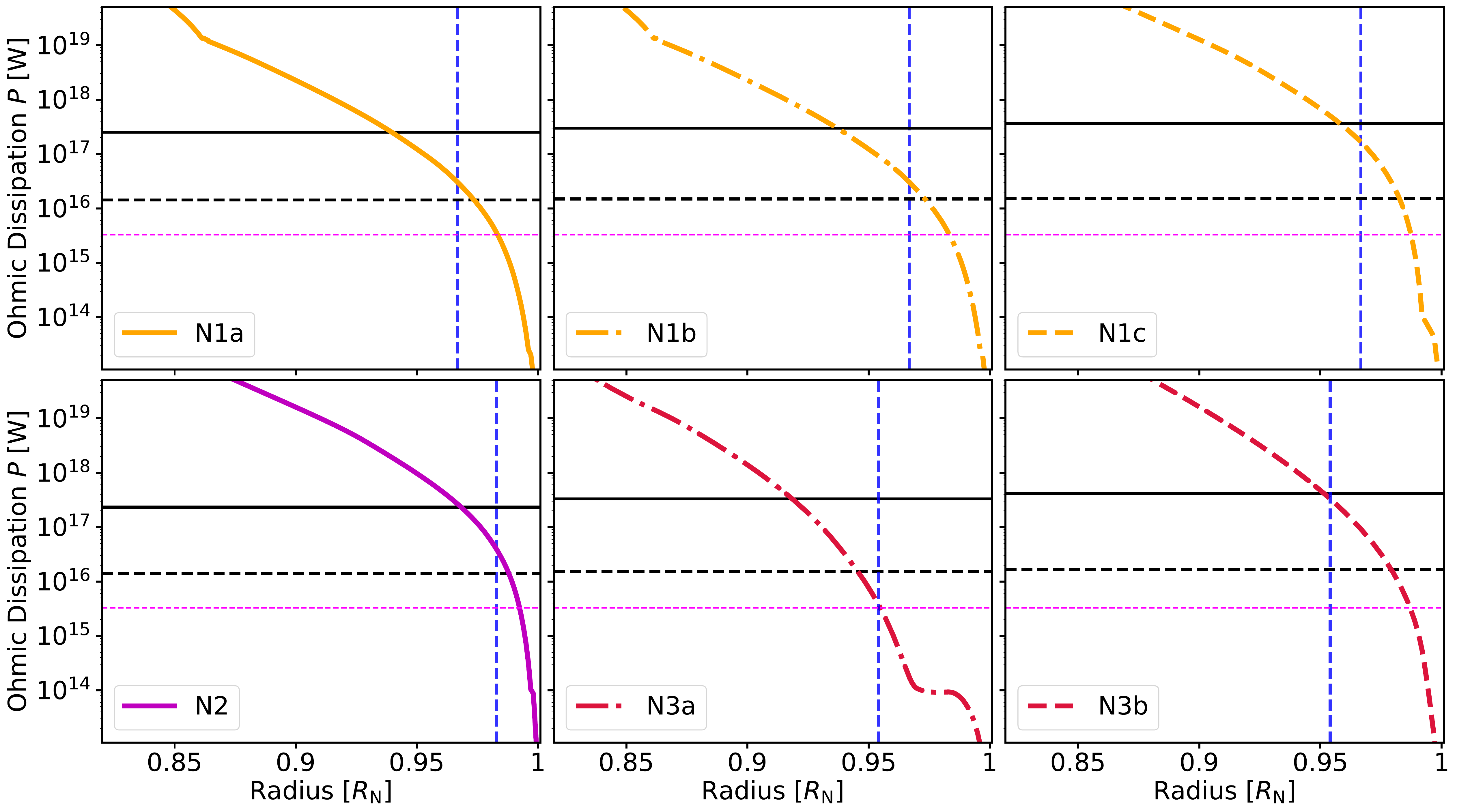}
    \caption{Same as Figure \ref{fig:u_ohm}, but for Neptune. The energy flux constraint limits the wind penetration to $0.95 - 0.98 R_{\mathrm{\scriptscriptstyle{N}}}$, depending on the model. The entropy flux limit  constrains the penetration to depths between $0.92 - 0.97 R_{\mathrm{\scriptscriptstyle{N}}}$.}
    \label{fig:n_ohm}
\end{figure*}

In this study the heavy elements in Uranus and Neptune were solely represented by water. Clearly, this is an unrealistic  assumption as more complex compositions are expected \citep{hf}. 
Given that both planets consist of other elements, no doubt the electrical conductivity profiles would be affected.  Nevertheless, the depth at which the energy/entropy budget is overrun is rather insensitive to changes of a factor of a few in the electrical conductivity value (as can be seen in Figures \ref{fig:u_ohm} and \ref{fig:n_ohm}).

In addition, the exclusion of rocky material, which was present in the models by \citet{vazan} (U3 and U4), presumably has lead to a higher electrical conductivity  than that expected from a silicate--water mixture. 
However, since these models are very hot, we would still expect the electrical conductivity values at shallow regions in U3 and U4 to be greater than those in the adiabatic models. 
At the same time our calculations could  underestimate the effective diffusion coefficients of hydrogen and oxygen for pressures $\lesssim 10$ GPa, which  would result in a  higher electrical conductivity and  therefore higher Ohmic dissipation. 
This takes place  in the relevant region for both Uranus and Neptune, since this pressure regime is reached at depths below $\sim 0.85 R_{\mathrm{\scriptscriptstyle{U}}}$ and $\sim 0.9 R_{\mathrm{\scriptscriptstyle{N}}}$, far deeper than where the Ohmic dissipation limit is reached for both planets. 
Nevertheless, as mentioned before, we take the conservative approach of not extrapolating the diffusion coefficients to lower pressures, in spirit of taking an upper bound for the energy/entropy budget violation depth.


In addition, we have used a water EOS which is different from the EOS in the original papers of the interior structure models. Although, this probably has an insignificant effect on the final results, the inferred water abundances for each model might be different from the original ones. However, we find that also models with small water fractions still surpass the Ohmic dissipation limit at relatively shallow depths. 
This suggests that not much water 
is required for the constraint to be overrun (e.g., the average water abundance for model N3a above $0.95 R_{\mathrm{\scriptscriptstyle{N}}}$ is $\sim 10\%$ by mass). As for the hydrogen and helium EOS, the \citet{cms} EOS is partly the same as the EOS used in modelling the interior structure models by \citet{nadine} and \citet{vazan}, since they have also used the SCvH EOS \citep{scvh}, which is a part of the former for low density regimes (see Section \ref{sec:int} for details on the EOS by \citet{cms}).

We find that the  electrical conductivity converges towards a common range of values around $\sim 0.7-0.75 R_{\mathrm{\scriptscriptstyle{U,N}}}$ for both planets, where they lie between $\sim 2\times10^3 - 10^4$ Sm$^{-1}$ in Uranus and Neptune at  $ 0.7 R_{\mathrm{\scriptscriptstyle{U,N}}}$.
Note that, these values are similar to what is usually used for modelling the conductivity of "ices" at that depth; namely $2 \times 10^3$Sm$^{-1}$ \citep{nellis_ice, holme, stanley}. The saturation depth of the conductivity profile agrees with the prediction of \citet{holme}.

Comparing estimates for the electrical conductivity values at the dynamo generation region, and constraining the values from additional physical arguments would contribute towards refining our ionic conductivity prescription. This could constrain the temperature profiles and the composition gradients of interior structure models.
As mentioned before, semi-conducting hydrogen has negligible contribution to the total electrical conductivity in our models above $\sim 0.9 R_{\mathrm{\scriptscriptstyle{U,N}}}$.
This conclusion is reached without even calculating the reduction to the electronic contribution (e.g. presence of helium and water in the mixture as extra scatterers and dissociation of H$_2$).

An important caveat when considering the total Ohmic dissipation is that, as $R_\textrm{m}$ exceeds unity, the ambiguity associated with the behaviour of the magnetic field and its interaction with the flow becomes more significant. 
The magnetic evolution becomes highly non-linear with depth, and it is unclear how deep one can use the current density approximations for computing the total Ohmic dissipation with confidence. Naturally, $R_\textrm{m}$ does not have the same value in all the points on a spherical shell. 
Thus, the $R_\textrm{m}$ calculation using the rms wind velocities $\langle U_\varphi \rangle$ should be taken with caution, since at some radii, a large portion of a thin shell in the planet will have $R_\textrm{m} < 1$, whereas the other part will have $R_\textrm{m} > 1$. Moreover,  uncertainties in the composition of the outer layers of the planets, and the disregard of atmospheric effects would eventually lead to misrepresentation of the value $R_\textrm{m}$. We note that the $R_\textrm{m}$ values for each model are typically within $\pm 3\% R_{\mathrm{\scriptscriptstyle{U,N}}}$ around where the models pass the entropy flux limit.

As noted by \citet{wicht}, the equations for the Ohmic limits ignore helium segregation, convective mixing of material, and planetary shrinking. 
These may well be negligible for our purposes, but the most important assumption is that the convection maintains an adiabat throughout the convective region of the planet. This is the case for half of the models that we adopt, the other half having a double diffusive convection \citep{podolak} or some mixture of convection/conduction \citep{vazan} instead. Moreover, the heat flux constraint comes with the additional requirement that the adiabatic cooling of the planet cancel out the dissipative heating at each radius. Nevertheless, most of the models also cross the less stringent entropy flux limit at depths consistent with the gravity estimations, especially in Neptune.

For the sake of simplicity, we assume that the zonal winds maintain their surface velocities with depth, in order to mimic a deep-seated wind profile. 
A more realistic model in these planets  resembles models where the winds decay with depth \citep{duer}. 
Including various decay modes, however, would have significantly complicated our analysis since it would have introduced additional free parameters related to the decay profile.  Clearly, further investigations accounting for different wind profiles as well as rotation rates are required and we hope to address this in future research.

In addition, the assumption that the winds penetrate inside along cylinders parallel to the rotation axis is very common when considering fast rotating planets like Uranus and Neptune. An interesting alternative could be a  penetration profile with latitudinal dependence, e.g. along radial lines at certain latitudes like that proposed for the Sun \citep{solar}. 
In any case, since gravity harmonic constraints are based on cylindrical behaviour as well, using the same penetration model makes these two phenomenologically independent constraints comparable.

\section{Conclusions}
\label{sec:con}
We provide constraints on the maximum penetration depths of zonal winds in Uranus and Neptune using the induced Ohmic dissipation due to the interaction of the winds and the planetary magnetic fields. 
We develop a method for calculating electrical conductivity profiles of ionically conducting hydrogen--helium--water mixtures under planetary conditions, using results from \textit{ab initio} simulations. 
Applying this prescription to various interior structure models suggests that deep-seated winds on Uranus and Neptune are unlikely. 
Our estimates are consistent with other maximum penetration depth estimates based on the dynamical contribution of the zonal winds on the spherical gravity harmonic $J_4$ \citepalias{kaspi}, which give an e-folding decrease depth of zonal wind strength around $0.95 R_{\mathrm{\scriptscriptstyle{U,N}}}$ for both planets. 

Using the energy flux constraints, we find that the maximum penetration depth for Uranus is between $0.93-0.97 R_{\mathrm{\scriptscriptstyle{U}}}$ and for Neptune between $0.95-0.98 R_{\mathrm{\scriptscriptstyle{N}}}$. Considering the entropy flux instead, this limit becomes $0.90-0.95 R_{\mathrm{\scriptscriptstyle{U}}}$ for Uranus and $0.92-0.97 R_{\mathrm{\scriptscriptstyle{N}}}$ for Neptune. 
Thus, our research suggests that for zonal winds penetrating along cylinders parallel to the rotation axis of the planets, the total induced Ohmic dissipation would be excessive if the wind velocities do not decay significantly with depth.

It should be noted that our results on the electrical conductivity corresponds to the case where the heavy elements are represented by pure-water. It is clear that using more representative compositions for the atmospheres of Uranus and Neptune are required. This in turn, can lead to different values for the electrical conductivity calculation. 
In addition, further investigations of the interplay between the wind profiles, rotation rates, compositions, and electrical conductivity are required. 

More accurate electrical conductivity estimates would provide important information that can be used to better understand the dynamo generation mechanism, the Ohmic dissipation associated with the dynamo, and subsequently, the convective as well as the thermal behaviour of the planets.  We therefore stress the importance of electrical conductivity estimates for Uranus and Neptune and their implications to better understand their internal structures and compositions.

Our method could be further refined with better constrained structure models of Uranus and Neptune.
However, it is unlikely that new data will be available on the interiors of ice giants without a new mission(s). Nevertheless, there are probably still unexplored areas for constraining contemporary structure models from various angles. 
In the meantime, the strong interplay of various phenomena like heat transfer, mixing, magnetic field generation, and fluid flow inside these planets provide a challenging, but fruitful playground for theoretical predictions. 

\section*{Acknowledgements}

DS thanks H.~Lascombes de Laroussilhe and C.~Valletta for valuable discussions. We thank O.~Shah and S.~Müller for providing us with the reliable EOS data. FS thanks Burkhard Militzer for allowing to reuse some simulations data for the water-hydrogen mixtures.
RH acknowledges support from the  Swiss National Science Foundation
(SNSF) via grant 200020\_188460 and fruitful discussions with D.~Stevenson and Y.~Kaspi.  
DS  acknowledges the help of mediocre quality coffee for conducting the research.

\section*{Data Availability}
For interior structure models:
The data were provided by [Morris Podolak, Allona Vazan] under licence / by permission. Data will be shared on request to the corresponding author with permission of [Morris Podolak, Allona Vazan].

For water equations of state:
The data were provided by [Oliver Shah] under licence / by permission. Data will be shared on request to the corresponding author with permission of [Oliver Shah].

For data in Figure 2 and D1:
The data were accessed from [CEA, via Francois Soubiran]. Data will be shared on request to the corresponding author with permission of [CEA, via Francois Soubiran].




\bibliographystyle{mnras}
\bibliography{example} 



\newpage

\appendix
\section{Current Density Estimation}
\label{sec:j}

Here we go through the derivation of Eq. (\ref{eq:j}) following \citet{liu_phd}.
Expressing the electrical field as a potential gradient: $\mathbf{E} = -\nabla \phi$, we can write down the solenoidality of the current density as:
\begin{align*}
\nabla \cdot \mathbf{j} =&\frac{1}{r^2} \frac{\partial}{\partial r}\left(r^2 \sigma(r)  \left(-(\nabla \phi)_r + (\mathbf{U}_\varphi \times \mathbf{B}_P)_r + (\mathbf{U}_\varphi \times \mathbf{B}_T)_r \right) \right) \nonumber \\ 
+& \frac{\sigma(r)}{r \sin{\theta}}\left[ \frac{\partial}{\partial \theta} \left(\sin{\theta}  \left(-(\nabla \phi)_\theta + (\mathbf{U}_\varphi \times \mathbf{B}_P)_\theta \right) \right) - \frac{\partial (\nabla \phi)_\varphi}{\partial \varphi}\right] \nonumber \\
=& 0.
\label{eq:sol}
\end{align*}
The dominant term here is the one concerning the derivative of the electrical conductivity. Therefore,
\begin{equation*}
0 \approx \partial_r \sigma(r)  \left(-\partial_r\phi + (\mathbf{U}_\varphi \times \mathbf{B}_P)_r + (\mathbf{U}_\varphi \times \mathbf{B}_T)_r \right),
\end{equation*}
implying that the terms inside the brackets cancel out. This is also motivated by the fact that current density in the radial direction $j_r$ is suppressed due to the spherically symmetric conductivity profile, confining the current to move along surfaces of similar electrical conductivity:
\begin{equation*}
0 \approx j_r = \sigma(r)  \left(-\partial_r\phi + (\mathbf{U}_\varphi \times \mathbf{B}_P)_r + (\mathbf{U}_\varphi \times \mathbf{B}_T)_r \right).
\end{equation*}
Integrating the term in the brackets along the radial direction, we get an expression for the electrical potential,
\begin{equation*}
\phi = \int\limits_r^R (\mathbf{U}_\varphi \times \mathbf{B}_P + \mathbf{U}_\varphi \times \mathbf{B}_T)_r dr' + K(\theta,\varphi).
\end{equation*}
Using this to get the electrical field in the $\theta$ direction and then plugging it in Eq. (\ref{eq:j}) to get $j_\theta$ (note that $\mathbf{U}_\varphi \times \mathbf{B}_T$ only has a radial component), we reach:
\begin{align*}
j_\theta \approx  \frac{\sigma(r)}{r}\Biggl(\frac{\partial}{\partial \theta} \int_r^R (\mathbf{U}_\varphi \times \mathbf{B}_P + \mathbf{U}_\varphi \times& \mathbf{B}_T)_r dr' \\
& + r  (\mathbf{U}_\varphi \times \mathbf{B}_P)_\theta + \frac{\partial K}{\partial \theta}\Biggr) 
\end{align*}
and analogously for $j_\varphi$:
\begin{equation*}
j_\varphi \approx \frac{\sigma(r)}{r}\Biggl(\frac{\partial}{\partial \varphi} \int_r^R (\mathbf{U}_\varphi \times \mathbf{B}_P + \mathbf{U}_\varphi \times \mathbf{B}_T)_r dr'+ \frac{\partial K}{\partial \varphi}\Biggr) 
\end{equation*}
As noted in Section \ref{sec:ohm}, $j_\phi$ has a negligible contribution to the total Ohmic dissipation compared to that of $j_\theta$. Thus, we go on with the calculation for only $j_\theta$. The goal is to bound the integration constant $\partial_\theta K$ in the case that it is negative (since positive constants would increase the Ohmic dissipation even more). Suppose the terms inside the brackets cancel out at $r_0$. Since the constant is independent of radial components, they will not cancel each other at $r_0 + l$, as the functions inside the brackets are monotonous along $r$ in the regions of interest.
We approximate the Ohmic dissipation at a shell of thickness $l$ at $r_0$, which has to be smaller than the heat flux (in \citet{liu_phd} this is taken as the surface luminosity):
\begin{equation*}
P_l \approx \frac{4\pi r{_0}^2 l}{\sigma}j_\theta^2 \lesssim \mathcal{E}_Q.
\end{equation*}
Thus, we can safely say:
\begin{align*}
&\left|\frac{\partial}{\partial \theta} \int_{r_0}^R (\mathbf{U}_\varphi \times \mathbf{B}_P + \mathbf{U}_\varphi \times \mathbf{B}_T)_r dr' + r_0  (\mathbf{U}_\varphi \times \mathbf{B}_P)_\theta + \frac{\partial K}{\partial \theta}\right| \\
&\leq \sqrt{\frac{\mathcal{E}_Q}{4 \pi \sigma l}}.
\end{align*}
Across this shell, the $\partial_i K$ do not change, whereas the other terms vary by $l/r_0$, to the first order. Thus, we can write (also analogously for the second equation above):
\begin{align}
&\left|\frac{\partial}{\partial \theta} \int_{r_0}^R (\mathbf{U}_\varphi \times \mathbf{B}_P + \mathbf{U}_\varphi \times \mathbf{B}_T)_r dr' + r_0  (\mathbf{U}_\varphi \times \mathbf{B}_P)_\theta\right| \nonumber \\ 
&\leq \frac{r_0}{l} \sqrt{\frac{\mathcal{E}_Q}{4 \pi \sigma l}}.
\label{eq:japp}
\end{align}
Since $\partial_\theta K$  is supposed to cancel the other terms at $r_0$, this is equivalent to:
\begin{equation*}
\left|\frac{\partial K}{\partial_\theta}\right| \leq \frac{r_0}{l} \sqrt{\frac{\mathcal{E}_Q}{4 \pi \sigma l}}
\end{equation*}
Evaluating these bounds where the ionic conductivity starts slowly to saturate in our estimates, for example around $\sim \!5000$ Sm$^{-1}$ (10000 Sm$^{-1}$) at $r \sim \!0.7 R_{\mathrm{\scriptscriptstyle{U}}} (R_{\mathrm{\scriptscriptstyle{N}}})$ for model U5c (N3b) and taking the thickness of the shell as the scale height of magnetic diffusivity, i.e.:
\begin{equation*}
l = H_\eta= \frac{\eta}{\partial_r \eta} = \frac{-\sigma}{\partial_r \sigma} \approx 2 \times 10^6 \; \mathrm{m}, \;\;\;\;\; \mathrm{at} \;\, r = 0.7 R_{\mathrm{\scriptscriptstyle{U,N}}} \;,
\end{equation*}
we calculate:
\begin{equation*}
\frac{r_0}{l} \sqrt{\frac{\mathcal{E}_Q}{4 \pi \sigma l}} \lesssim 2 \times 10^3 \;\mathrm{Tm^2s^{-1}} \; \mathrm{for ~ Uranus}\; \mathrm{and} \; \mathrm{Neptune.} \\
\end{equation*}
Note that, doing the same calculation with the entropy flux limit $\mathcal{E}_s$, the estimates increase by a factor of $\sqrt{\mathcal{E}_S/\mathcal{E}_Q} \sim 3$. Since $K(\theta,\varphi)$ is not a function of radius, this limit is also valid for shallow regions.
Taking a look at the outermost region, we expect the l.h.s of Eq. (\ref{eq:japp}) to be of order $\left|r \mathbf{U}_\varphi \mathbf{B}_P  \right|$. Using the average value of magnetic field strength (0.23 G and 0.14 G at the equator, for Uranus and Neptune, respectively) and the averaged wind speed over the surface, we determine:
\begin{equation*}
\left|r \mathbf{U}_\varphi \mathbf{B}_P  \right| \approx 7 \times 10^4 \;\mathrm{Tm^2s^{-1}} \; \mathrm{for ~ Uranus}\; \mathrm{and} \; \mathrm{Neptune.}
\end{equation*}
Thus, more than an order of magnitude greater than the estimates for $\partial_\theta K$. Finally, noting that the toroidal magnetic field strength $|\mathbf{B}_T|$ is of the order of $R_\textrm{m}|\mathbf{B}_P|$, we reach the current density profiles  in Eq. (\ref{eq:jthe}), (\ref{eq:jphi}).
Clearly, this is an over-simplification of the dynamics involved in the evolution and the behaviour of the magnetic field and its interaction with the zonal flow. Although, there is no simple way of determining the true poloidal-toroidal coupling, its evolution and the true behaviour of the integration constant above, the formulation leading to  Eq. (\ref{eq:jthe}), (\ref{eq:jphi}) is still a good indicator of the current density in the shallow layers of the planet, where the behaviour of the system is less unpredictable.

\section{Total Ohmic Dissipation Term}
\label{sec:biggest_boy}

For clarity, we explicitly write out the total Ohmic dissipation $P_{\mathrm{tot}}$. We separate $P_{\mathrm{tot}}$ as the dissipation arising due to currents in $\theta$ and $\varphi$ directions: $P_{\mathrm{tot}} = P_\theta + P_\varphi$. Thus, it follows that:

\begin{equation*}
P_\theta = \int\limits_r^R\int\limits_0^\pi\int\limits_0^{2\pi} \frac{\left( j_{\theta,1}(r',\theta,\varphi) + j_{\theta,2}(r',\theta,\varphi)\right)^2}{\sigma(r')}  \; r'^2\sin{\theta}dr'd\theta d\varphi, 
\end{equation*}
where
\begin{align*}
j_{\theta,1}  =  \frac{\sigma(r')}{r'}  \sum^\infty_{l=1}&\sum^l_{m=0} R^{l+2}
(g_l^m \cos{m\varphi} + h_l^m \sin{m\varphi}) ~\times \\
&\Biggl(\frac{\partial^2 P_l^m (\cos\theta)}{\partial \theta^2} \int\limits_{r'}^R \dfrac{v_\varphi\!\left(\arcsin{\left(\frac{r'' \sin{\theta}}{R}\right)}\right)}{r''^{l+2}}dr'' \\
&+ \frac{\partial P_l^m (\cos\theta)}{\partial \theta}  \int\limits_{r'}^R \dfrac{dr''}{r''^{l+2}} \dfrac{\partial v_\varphi \left(\arcsin{\left(\frac{r'' \sin{\theta}}{R}\right)}\right)}{\partial \theta}
\Biggr)
\end{align*}
and
\begin{align*}
j_{\theta,2} = \sigma(r')  \sum^\infty_{l=1}\sum^l_{m=0} (l+1)\left(\frac{R}{r'}\right)^{l+2}
&(g_l^m \cos{m\varphi} + h_l^m \sin{m\varphi}) \times \\ 
&P_l^m(\cos\theta)  v_\varphi\!\left(\arcsin{\left(\frac{r' \sin{\theta}}{R}\right)}\right).
\end{align*}
Then also,
\begin{equation*}
P_\varphi = \int\limits_r^R\int\limits_0^\pi\int\limits_0^{2\pi}  \frac{j_\varphi^2(r',\theta,\varphi)}{\sigma(r') } \; r'^2 \sin{\theta}dr'd\theta d\varphi, 
\label{eq:p_phi}
\end{equation*}
where
\begin{align*}
j_\varphi = \frac{\sigma(r')}{r'}  \sum^\infty_{l=1}\sum^l_{m=1}
&R^{l+2}  m  (-g_l^m \sin{m\varphi} + h_l^m \cos{m\varphi}) \times \\
&\frac{\partial P_l^m(\cos\theta)}{\partial \theta} {\int\limits^R_{r'} \dfrac{v_\varphi\!\left(\arcsin{\left(\frac{r'' \sin{\theta}}{R}\right)}\right)}{r''^{l+2}} dr''}.
\end{align*}
We compute the total Ohmic dissipation numerically, using the magnetic field expansion up to $l=3$ (i.e. up to octopole).

\section{DC limit of the Drude Model}
\label{sec:electronic}
Starting from the modified version of the Drude model \citep{celliers}:
\begin{equation}
    \sigma(\omega)=\frac{n_{\textrm{\scriptsize e}}e^2\tau}{2 m_{\textrm{\scriptsize eff}}}\frac{1}{1-i\omega\tau}, 
\end{equation}
where $\sigma(\omega)$ is the electrical conductivity at frequency $\omega$, $n_{\textrm{\scriptsize e}}$ is the electron density, $e$ the electron charge, $\tau$ the dissipation time, and $m_{\textrm{\scriptsize eff}}$ is an effective mass.
For a semi-conductor, the electron density is given by:
\begin{equation}
    n_{\textrm{\scriptsize e}}=2 \left ( \frac{m_{\textrm{\scriptsize eff}}k_\textrm{\scriptsize B}T}{2\pi\hbar^2}\right)^{3/2}f_{1/2}\left (-\frac{E_\textrm{\scriptsize g}}{2k_\textrm{\scriptsize B}T}\right) ,
\end{equation}
with $k_\textrm{\scriptsize B}$ being the Boltzmann constant, $\hbar$ the reduced Planck constant, $T$ the temperature, $E_\textrm{\scriptsize g}$ the energy gap, and the Fermi function:
\begin{equation} \label{eq:fermi}
    f_m(x)=\frac{2}{\sqrt{\pi}}\int_0^{+\infty}\textrm{d}y y^m \frac{1}{1+e^{y-x}}.
\end{equation}
Finally, for the dissipation time, within the Mott-Ioffe-Regel limit, we get $\tau=a/v_\textrm{\scriptsize e}$ with $a$ the interparticle distance and the velocity defined as \citep{celliers}:
\begin{equation}
    v_\textrm{\scriptsize e}^2=\dfrac{2k_\textrm{\scriptsize B}T}{m_{\textrm{\scriptsize eff}}}\dfrac{f_{3/2}\left (-\dfrac{E_\textrm{\scriptsize g}}{2k_\textrm{\scriptsize B}T}\right)}{f_{1/2}\left (-\dfrac{E_\textrm{\scriptsize g}}{2k_\textrm{\scriptsize B}T}\right)}.
\end{equation}
In the limit of low temperatures Eq. (\ref{eq:fermi}) can be approximated by:
\begin{equation} 
    f_m(x)\simeq\frac{2}{\sqrt{\pi}}\int_0^{+\infty}\textrm{d}y y^m e^{-y} e^{-E_\textrm{\scriptsize g}/2k_\textrm{\scriptsize B}T}=\frac{2}{\sqrt{\pi}}\Gamma_{m+1}e^{-E_\textrm{\scriptsize g}/2k_\textrm{\scriptsize B}T},
\end{equation}
via the $\Gamma$-function.
After some algebra, we obtain for the electrical conductivity:
\begin{equation}\label{eq:condsemicond}
    \sigma(\omega)=\frac{4\pi^{3/2}}{\sqrt{3}}\frac{e^2m_{\textrm{\scriptsize eff}}k_\textrm{\scriptsize B}Ta}{h^3}e^{-E_\textrm{\scriptsize g}/2k_\textrm{\tiny B}T}\frac{1}{1-i\omega\tau}.
\end{equation}
In the DC limit, this term becomes Eq. (\ref{eq:elec_cond}):
\begin{equation}
    \sigma(0) = \sigma_0 e^{-E_\textrm{\scriptsize g}/2k_\textrm{\scriptsize B}T}
\end{equation}
\section{Pressure Functions for Dissociation Fractions}
\label{sec:frac}
\begin{figure*}
    \centering
    \includegraphics[width = 2\columnwidth]{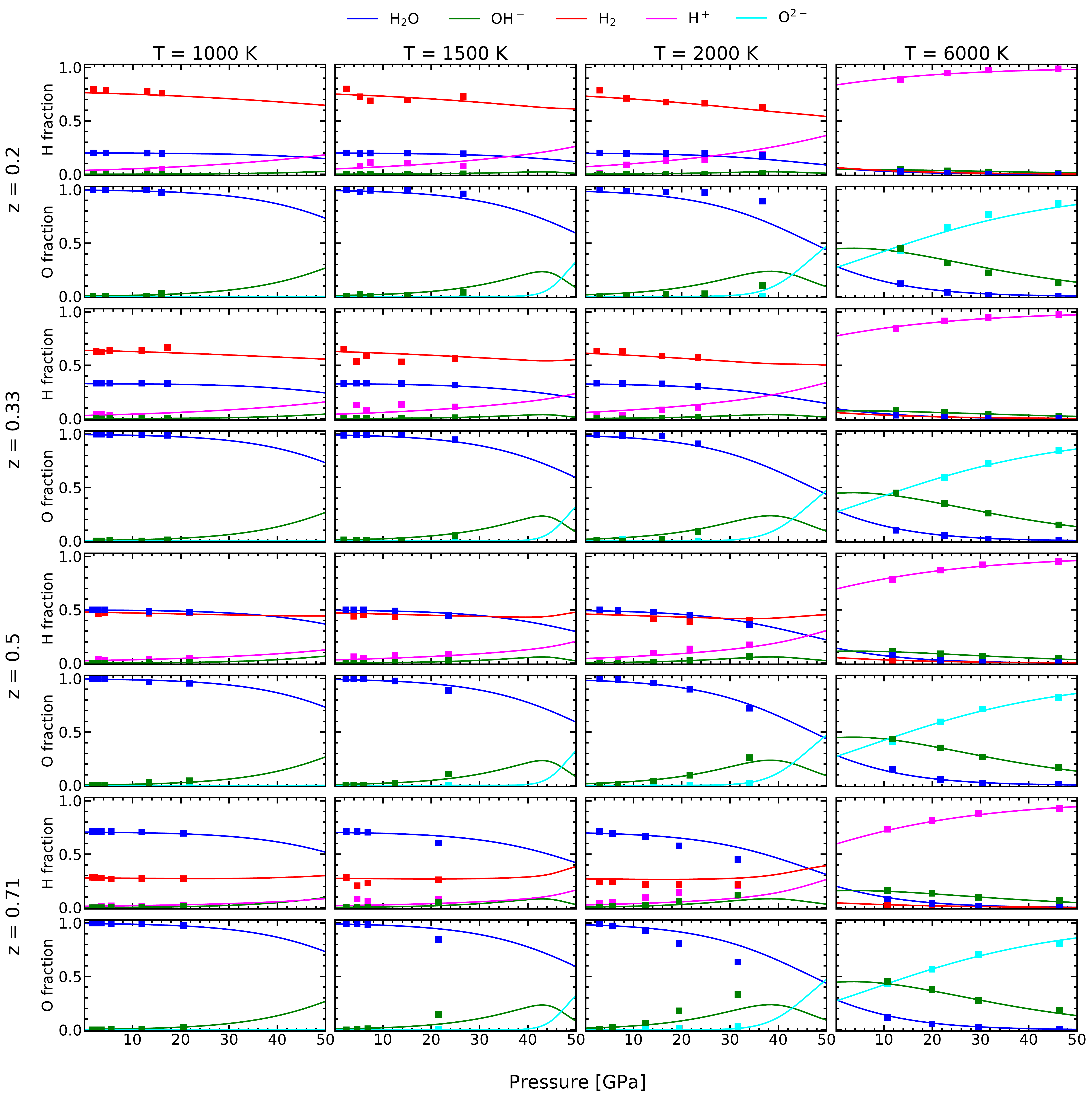}
    \caption{Fraction of H in species H$^+$, H$_2$, H$_2$O, OH$^-$ and fraction of O in species H$_2$O, OH$^-$ and O$^{2-}$ in a H$_2$--H$_2$O mixture. Every column indicates a different temperature regime and z denotes the molecular ratio of H$_2$O to H$_2$ (i.e $z  = n_{\mathrm{H_2O}}/(n_\mathrm{H_2O}+n_\mathrm{H_2})$). Note that, with this definition, the number of oxygen atoms is: $n_\mathrm{O} = z  n_\mathrm{H} / 2$. The data points are from \citet{francois_diss}.}
    \label{fig:frac}
\end{figure*}

The pressure functions that we use to calculate the dissociation fractions of molecules are listed below:
\begin{align*}
&P_{\textrm{\scriptsize H$_2$O}} = 75.85 - 1.44 \times 10^{-2} \left(\frac{T}{1 \textrm{K}}\right)  \nonumber \\
&P_{\textrm{\scriptsize H$_2$}} = 110.8 - 2.53 \times 10^{-2} \left(\frac{T}{1 \textrm{K}}\right)  \nonumber \\
&P_{\textrm{\scriptsize OH$^{-}$}} = 59.06 - 8.07 \times 10^{-3} \left(\frac{T}{1 \textrm{K}}\right)  \\
&\delta P_{\textrm{\scriptsize H$_2$O}} = 22.84 - 6\times 10^{-6} \left(\frac{T}{1 \textrm{K}}\right) \nonumber \\
& \delta P_{\textrm{\scriptsize H$_2$}} = 61.2 - 4.91 \times 10^{-3} \left(\frac{T}{1 \textrm{K}}\right) \nonumber \\
& \delta P_{\textrm{\scriptsize OH$^{-}$}} =  -8.7 + 8.4 \times 10^{-3} \left(\frac{T}{1 \textrm{K}}\right) \nonumber 
\end{align*}
We compute the number fraction of H and O in every molecule species using the dissociation fraction (Eq. (\ref{eq:cons})) and the conservation rule of total fraction of atoms (Eq. (\ref{eq:diss})).
As mentioned in the text, the number fractions of H and O in each molecule for varying temperature and water ratio is shown in Figure \ref{fig:frac} as a function of pressure.

Note that, we only take into consideration the number fractions in a hydrogen--water mixture and add helium fraction into the equations afterwards. There is no apparent reason for helium to have an effect on the number fractions of H and O in various molecule species in an ideal mixture.
Furthermore, the number fraction of atoms $\alpha$ in species $\beta$; $\xi_{\alpha,\beta}$ is defined as:
\begin{align*}
&\xi_{\textrm{\tiny O},\textrm{\tiny H$_2$O}} = \epsilon_{\textrm{\tiny H$_2$O}} \\
&\xi_{\textrm{\tiny H},\textrm{\tiny H$_2$O}} = z  \epsilon_{\textrm{\tiny H$_2$O}} \\
&\xi_{\textrm{\tiny O},\textrm{\tiny OH$^-$}} = (1 - \epsilon_{\textrm{\tiny H$_2$O}})  \epsilon_{\textrm{\tiny OH$^-$}} \\
&\xi_{\textrm{\tiny H},\textrm{\tiny OH$^-$}} = z  (1 - \epsilon_{\textrm{\tiny H$_2$O}})  \epsilon_{\textrm{\tiny OH$^-$}}/2 \\
&\xi_{\textrm{\tiny O},\textrm{\tiny O$^{2-}$}} = 1 - \xi_{\textrm{\tiny O},\textrm{\tiny H$_2$O}} - \xi_{\textrm{\tiny O},\textrm{\tiny OH$^-$}} \\
&\xi_{\textrm{\tiny H},\textrm{\tiny H$_2$}} ~~ = (1 - \xi_{\textrm{\tiny H},\textrm{\tiny H$_2$O}} - \xi_{\textrm{\tiny H},\textrm{\tiny OH$^-$}})  \epsilon_{\textrm{\tiny H$_2$}} \\
&\xi_{\textrm{\tiny H},\textrm{\tiny H$^{\Theta +}$}} = 1 - \xi_{\textrm{\tiny O},\textrm{\tiny H$_2$O}} - \xi_{\textrm{\tiny H},\textrm{\tiny OH$^-$}} - \xi_{\textrm{\tiny H},\textrm{\tiny H$_2$}}
\end{align*}
where $z$ is the molecular mixing ratio defined in Eq. (\ref{eq:z}).

\section{Diffusion Coefficients}
\label{sec:coeff}
For diffusion coefficients in cm$^2$s$^{-1}$, pressures in GPa and temperatures in K, the parameters values in Eq. (\ref{eq:diff_h})-(\ref{eq:diff_o}) are:
\begin{eqnarray}
    P_{0,\mathrm{H}}&=&0.00312552 \nonumber\\
    a_{\mathrm{H}}&=&-0.74285462 \nonumber\\
    b_{\mathrm{H}}&=& 0.00551773 \nonumber\\
    P_{0,\mathrm{O}}&=&0.00075243 \nonumber\\
    a_{\mathrm{O}}&=&-1.35077937 \nonumber\\
    b_{\mathrm{O}}&=&0.04149280 \nonumber
 \end{eqnarray}

The parameters of the modifying functions in Eq. (\ref{eq:diff_mod}) are then:
\begin{eqnarray}
    \alpha&=&0.6419031 \nonumber\\
    \beta&=&0.0075469 \nonumber\\
    \gamma&=& 0.6274056 \nonumber\\
    \delta&=&0.0110279 \nonumber
\end{eqnarray}



\bsp	
\label{lastpage}
\end{document}